\documentclass[journal]{IEEEtran}
\usepackage{graphicx}  
\usepackage{amsmath}
\usepackage{amssymb}
\usepackage{hyperref}
\usepackage{url}

\newcommand{\nclus}{\ensuremath{n_\text{clus}}}
\newcommand{\ncomp}{\ensuremath{n_\text{comp}}}
\newcommand{\nanode}{\ensuremath{n_\text{anode}}}
\newcommand{\nedge}{\ensuremath{n_\text{edge}}}
\newcommand{\nvox}{\ensuremath{n_\text{vox}}}
\newcommand{\nbins}{\ensuremath{n_\text{bins}}}

\title{Data-driven performance optimization of gamma spectrometers with many channels}
\author{
    Jayson~R.~Vavrek, Hannah~S.~Parrilla, Gabriel~Aversano, Mark~S.~Bandstra, Micah~Folsom, Daniel~Hellfeld
    \thanks{
        J.R.~Vavrek, H.S.~Parrilla, G.~Aversano, M.S.~Bandstra, M.~Folsom, and D.~Hellfeld are with the Nuclear Science Division, Lawrence Berkeley National Laboratory, Berkeley, CA, 94720 USA.
        The work presented in this paper was funded by the National Nuclear Security Administration of the Department of Energy, Office of International Nuclear Safeguards.
        This work was performed under the auspices of the U.S.\ Department of Energy by Lawrence Berkeley National Laboratory (LBNL) under Contract DE-AC02-05CH11231.
        The U.S.\ Government retains, and the publisher, by accepting the article for publication, acknowledges, that the U.S.\ Government retains a non-exclusive, paid-up, irrevocable, world-wide license to publish or reproduce the published form of this manuscript, or allow others to do so, for U.S.\ Government purposes.
    }
}
\date{\today}

\begin{document}

\maketitle

\begin{abstract}
    In gamma spectrometers with variable spectroscopic performance across many channels (e.g., many pixels or voxels), a tradeoff exists between including data from successively worse-performing readout channels and increasing efficiency.
    Brute-force calculation of the optimal set of included channels is exponentially infeasible as the number of channels grows, and approximate methods are required.
    In this work, we present a data-driven framework for attempting to find near-optimal sets of included detector channels.
    The framework leverages non-negative matrix factorization (NMF) to learn the behavior of gamma spectra across the detector, and clusters similarly-performing detector channels together.
    Performance comparisons are then made between spectra with channel clusters removed, which is more feasible than brute force.
    The framework is general and can be applied to arbitrary, user-defined performance metrics depending on the application.
    We apply this framework to optimizing gamma spectra measured by H3D M400 CdZnTe spectrometers, which exhibit variable performance across their crystal volumes.
    In particular, we show several examples optimizing various performance metrics for uranium and plutonium gamma spectra in nondestructive assay for nuclear safeguards, and explore trends in performance vs.\ parameters such as clustering algorithm type.
    We also compare the NMF+clustering pipeline to several non-machine-learning algorithms, including several greedy algorithms.
    Overall, we find that the NMF+clustering pipeline tends to find the best-performing set of detector voxels, significantly improving over the un-optimized spectra, but that a greedy accumulation of spectra segmented by detector depth can in some cases give similar performance improvements in much less computation time.
\end{abstract}

\section{Introduction}

Pixelated CdZnTe (CZT) gamma detectors have recently become an attractive technology for the non-destructive assay (NDA) of radiological materials.
CZT offers energy resolutions of ${\lesssim}1\%$ at $662$~keV, but operates at room temperature.
In particular, high-efficiency, large-volume CZT detector systems are commercially available from H3D, Inc.\ (Ann Arbor, MI, USA), and their M400 series of detectors is being evaluated by H3D, Inc., the International Atomic Energy Agency (IAEA), and the US National Laboratories to replace the HM-5 sodium iodide (NaI) handheld detector used for the majority of IAEA safeguards NDA measurements~\cite{iaea2022disp, dodane2023large, smith2024summary, streicher2016special}.

\begin{figure*}
    \centering
    \includegraphics[width=0.32\linewidth]{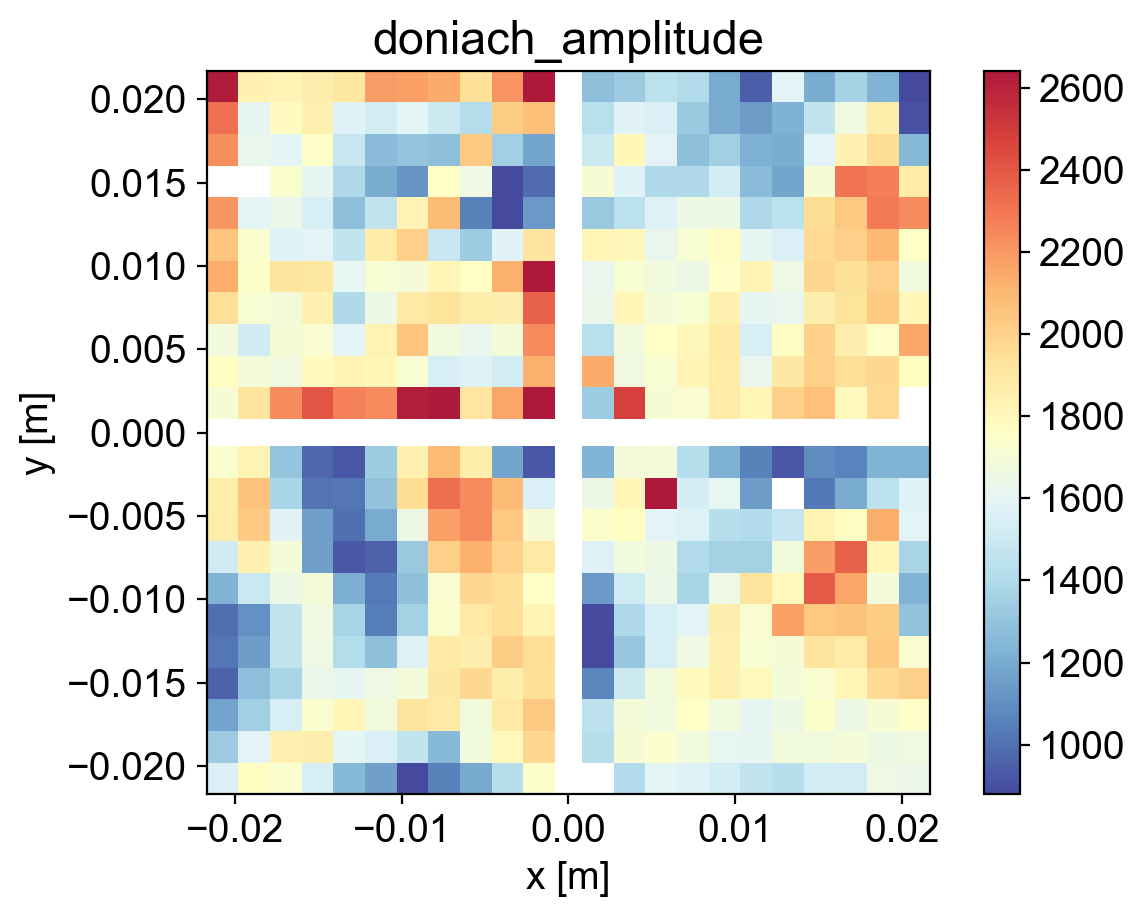}
    \includegraphics[width=0.32\linewidth]{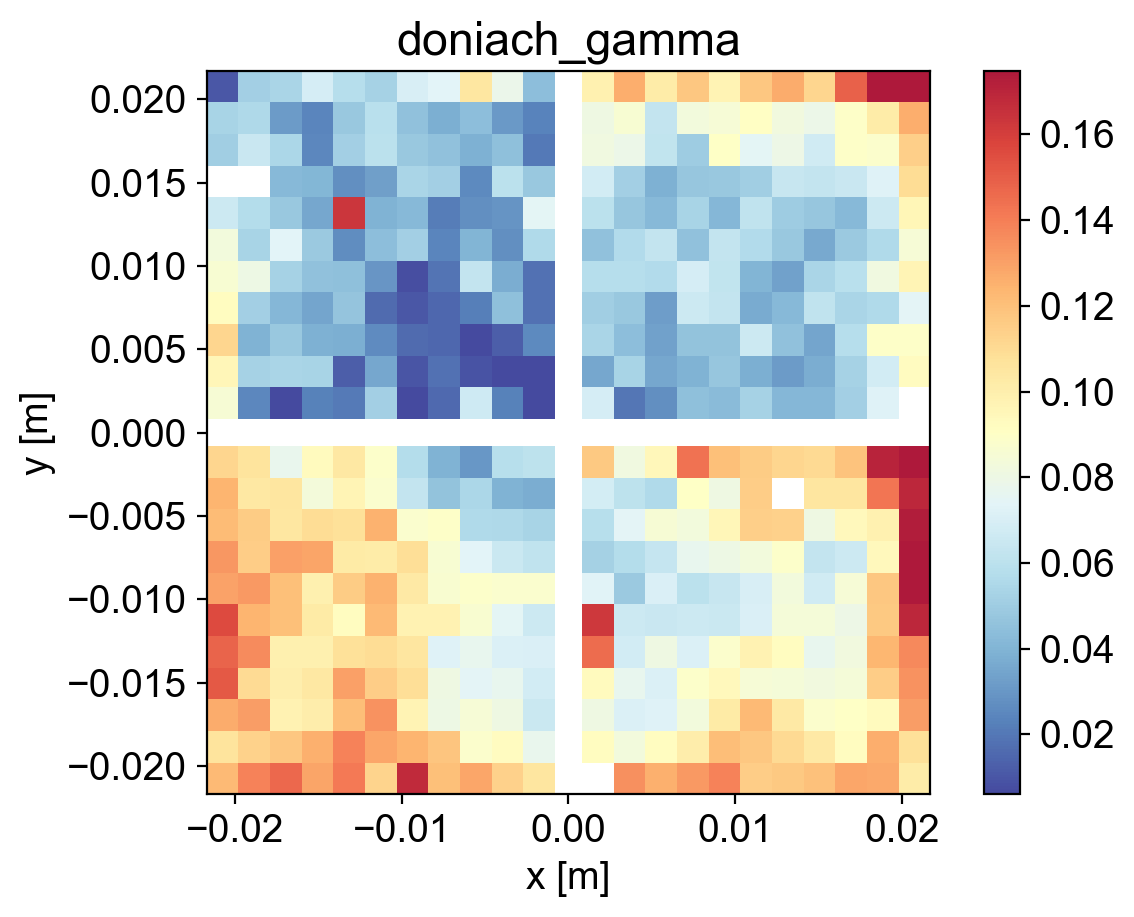}
    \includegraphics[width=0.32\linewidth]{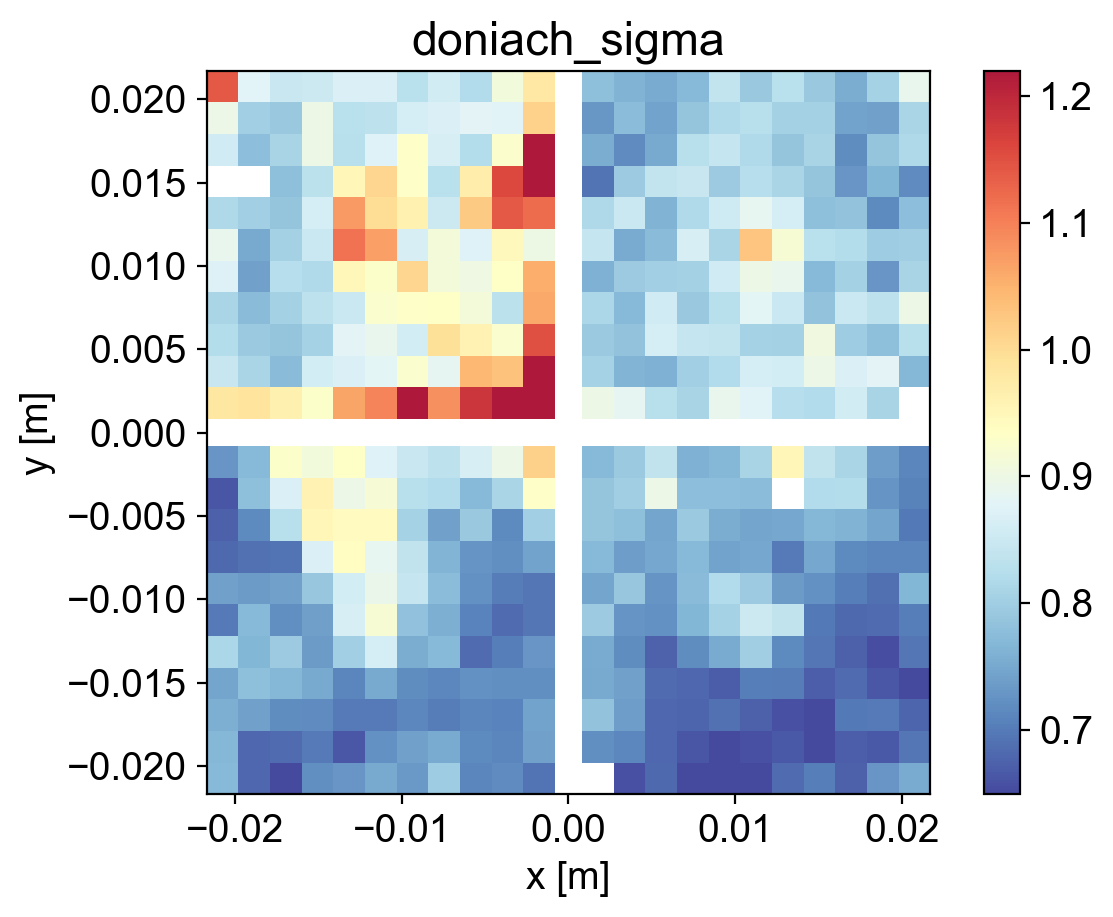}\\
    \includegraphics[width=0.32\linewidth]{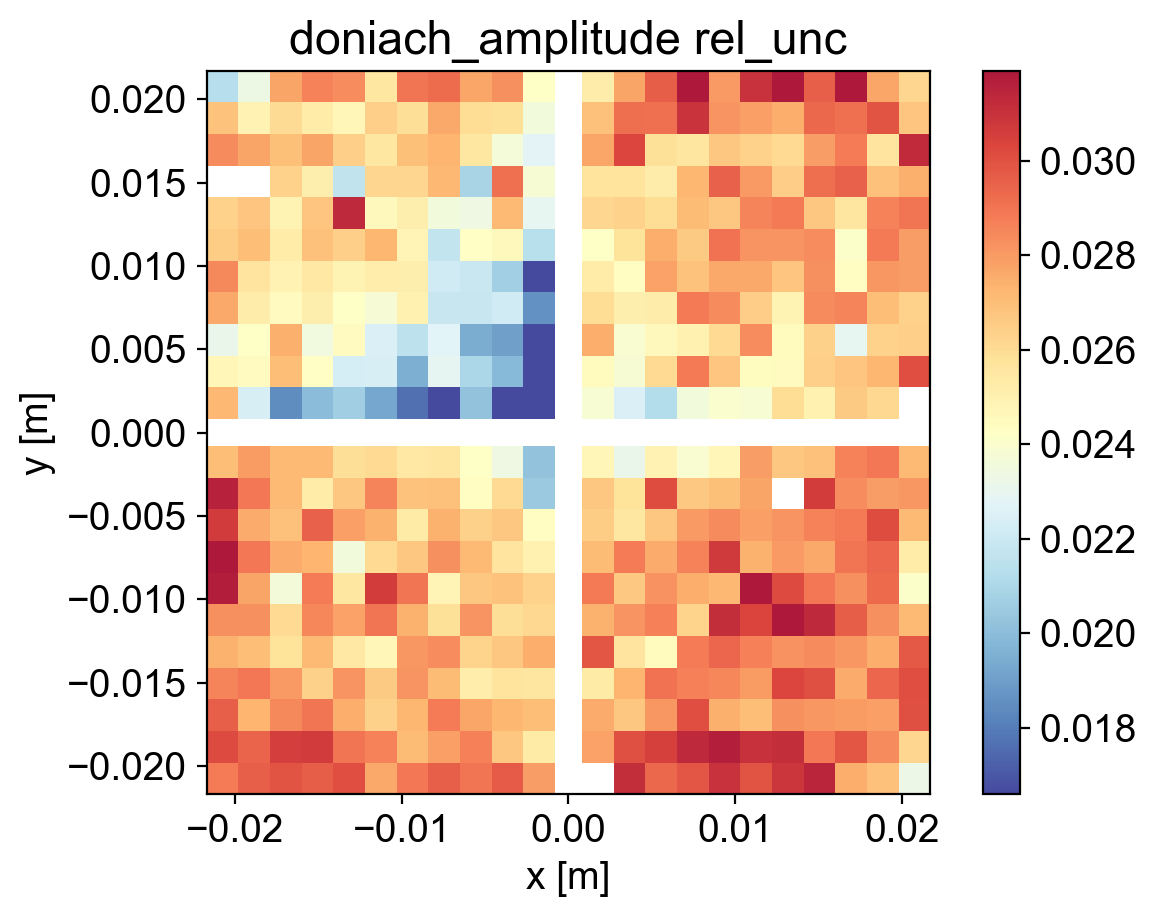}
    \includegraphics[width=0.32\linewidth]{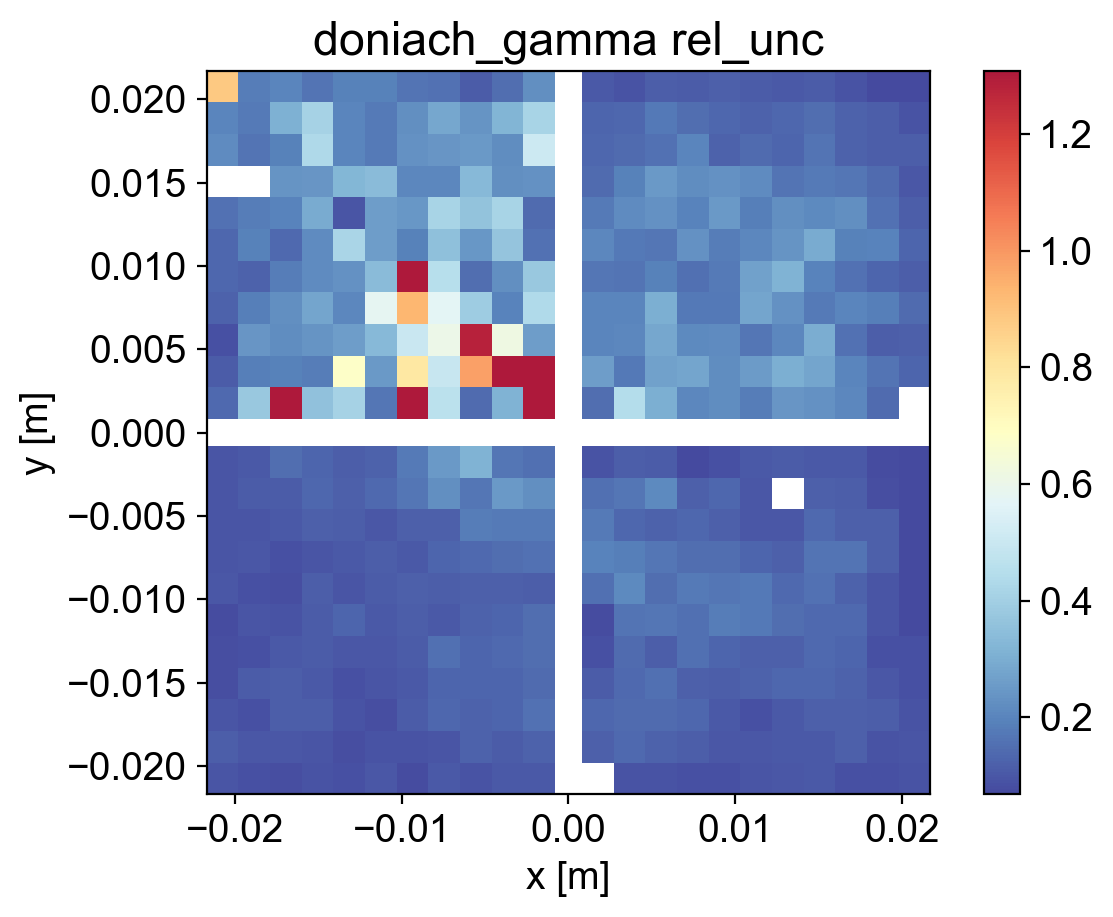}
    \includegraphics[width=0.32\linewidth]{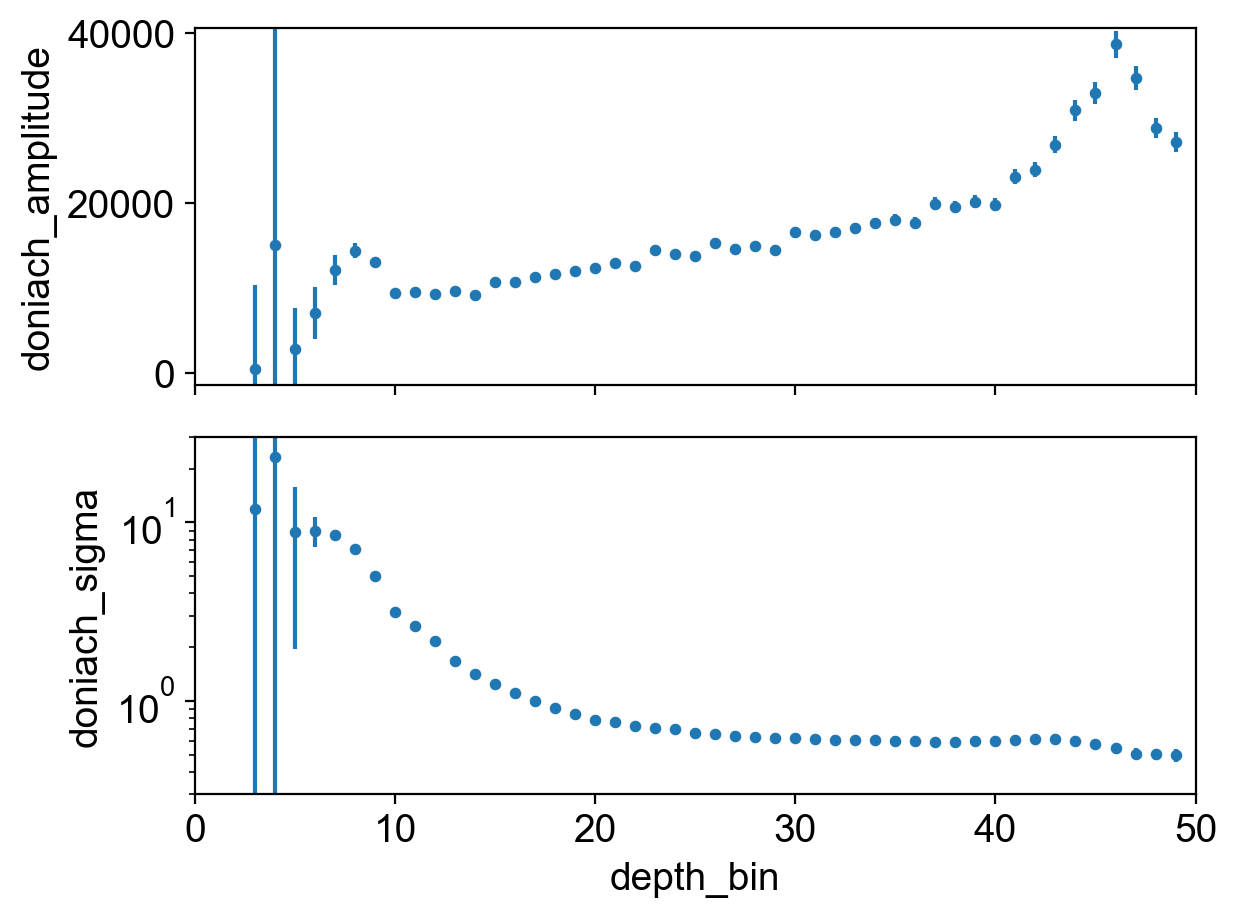}
    \caption{
        Variation of spectral performance over one M400 detector, quantified by Doniach peak fit parameters for spectra within each pixel or depth bin (Eq.~\ref{eq:doniach}).
        Top left: Doniach amplitude vs.\ pixel.
        Bottom left: Doniach amplitude relative uncertainty.
        Top center: Doniach~$\gamma$ (asymmetry).
        Bottom center: Doniach~$\gamma$ (asymmetry) relative uncertainty.
        Top right: Doniach~$\sigma$ (width).
        Bottom right: Doniach amplitude and $\sigma$ vs.\ depth bin.
    }
    \label{fig:fit-variations}
\end{figure*}

The pixelization of the M400 detector combined with depth-of-interaction estimation allows one to estimate the 3D position of gamma ray interactions within the detector crystal.
While this capability is typically used to enable Compton imaging of radiological sources~\cite{hellfeld2022quantitative, hecla2021polaris}, the 3D position information can also be used to evaluate and improve the spectroscopic performance of the detector.
For instance, after discretizing the depth dimension of the detector, individual voxels offer superior energy resolution ($0.65\%$ at 662~keV) compared to accumulating data from the entire ``bulk'' detector (${\sim}1\%$)~\cite{m400_spec_sheet}.
Other spectral performance metrics such as efficiency and peak tailing also vary across the detector volume---see Fig.~\ref{fig:fit-variations} of this work, Fig.~4 of Ref.~\cite{aversano2024unsupervised}, and Figs.~5--7 of Ref.~\cite{li1999spatial}.
In general, this creates a \textit{performance tradeoff} as spectra from each voxel are accumulated---using only the single best-performing voxel will sacrifice nearly all the detector efficiency and drastically increase measurement times, while using all voxels will maximize efficiency but include poorly-performing voxels.
For example tradeoff curves, see the ``greedy algorithm'' curves in \cite[Fig.~6]{aversano2024unsupervised}.

Between these two extremes, some combination of detector voxels will provide the optimal tradeoff between individual voxel performance and detector efficiency.
Exactly finding this optimal combination of voxels, however, is computationally infeasible---the M400 has $4$~CZT crystals, each pixelated to an $11 \times 11$ grid, and depth information can be discretized to (say) $50$~bins, resulting in $2^{24\,200} \simeq 10^{7285}$ possible voxel combinations.
But while an exact, brute force search is infeasible, the plots of Fig.~\ref{fig:fit-variations} suggest that there are spatial correlations between nearby voxels, and therefore that the detector may be divided into a small number of \textit{voxel clusters} with similar performance ($\nclus \simeq 2$--$7$).
An approximate search can then be feasibly performed over combinations of clusters, rather than combinations of individual voxels.

In this work, we present a data-driven framework for learning spectral correlations between detector voxels, building voxel clusters of similar spectral performance, comparing voxel cluster combinations in terms of user-defined performance metrics, and ultimately using the best voxel cluster combination to define a binary \textit{voxel mask} indicating which detector voxels to use for analysis.
We demonstrate the framework on several nuclear-safeguards-relevant datasets, aiming to optimize gamma spectra performance metrics that would feed into downstream NDA applications such as uranium enrichment calculations for samples of nuclear fuel materials perhaps $10$s of g to $10$s of kg~\cite{iaea2022itvs}.
The framework is however agnostic to the exact spectral performance metric and the detector architecture, and could in theory be applied to other highly-segmented gamma ray detectors with spatially-varying performance including germanium double-sided strip detectors (DSSDs)~\cite{amman2000three} such as the GeGI from PHDS Co.\ (Knoxville, TN, USA)~\cite{phdsco_gegi, jacomb2020spatially}, multi-crystal $4\pi$ gamma imagers such as those developed at Lawrence Berkeley National Laboratory (LBNL)~\cite{pavlovsky2019miniprism, vavrek2025demonstration}, or pixelated semiconductor arrays for energy-resolved positron emission tomography (PET) imaging~\cite{cates2015direct}.
Similarly, the framework could be applied to large physics experiments with many energy readout channels such as CUORE~\cite{gorla2012cuore, nutini2020cuore}, CUPID~\cite{armengaud2020cupid}, GRETA~\cite{vetter2000three}, or LEGEND~\cite{zsigmond2020legend}, ultra-high-resolution microcalorimeter arrays~\cite{hoover2009microcalorimeter, bacrania2009large, ullom2015review}, or more generally, to any dataset with generalized discrete ``regions'' of variable generalized ``performance'' that can be combined along some performance tradeoff curve.

This paper builds off of several previous works.
We leverage past expertise in applying non-negative matrix factorization (NMF)~\cite{lee1999learning, wang2012nonnegative} for learning patterns in gamma spectra~\cite{bilton2019non, bandstra2020modeling, bandstra2021correlations, bandstra2023background}.
Ref.~\cite{aversano2024unsupervised} was an earlier pipeline that clustered detector \textit{pixels} rather than voxels, and based the clustering on spatial variations in peak fit parameters rather than NMF weights as in the present work.
Ref.~\cite{aversano2023data} was an earlier proof-of-concept of the present workflow, but lacked a number of important features and analyses such as additional performance metrics, comparisons against non-machine-learning-based methods, and quantitative demonstrations on safeguards-relevant nuclides such as U and Pu.

The structure of this paper is as follows.
Section~\ref{sec:methods} introduces the various algorithm pipelines developed as well as the spectral performance metrics tested.
Section~\ref{sec:results} provides four example optimization problems, showing spectral improvements across various performance metrics and source spectra.
Section~\ref{sec:discussion} then provides additional discussion, including limitations of the present study to be addressed in future work, and considers opportunities for further operationalization of these algorithms and the generalizability of results to different detectors.

\section{Methods}\label{sec:methods}

Here we introduce four methods for clustering and removing low-performing detector channels: machine-learning-based clustering, heuristic clustering, random clustering, and greedy clustering.
While these methods may differ substantially in their behavior and performance, they each culminate in a binary voxel mask specifying which detector voxels to keep in order to optimize the given performance metric.

All four methods have been developed into the software package Spectral Peak Enhancement by Combining Trusted Response Elements via Machine Learning ({\tt spectre-ml}), which can be made available under either an academic/government/nonprofit license or a commercial license from the LBNL Intellectual Property Office~\cite{vavrek2023spectral}.

In the following sections, results labeled SPECTRE-ML are those requiring parameter sweeps---machine learning, heuristic, and random, but \textit{not} greedy, which are separately labeled.
We note that for all of the clustering methods, the channel clusters need not be spatially contiguous.

\subsection{ML-based clustering}

\begin{figure*}
    \centering
    \includegraphics[width=1.0\linewidth]{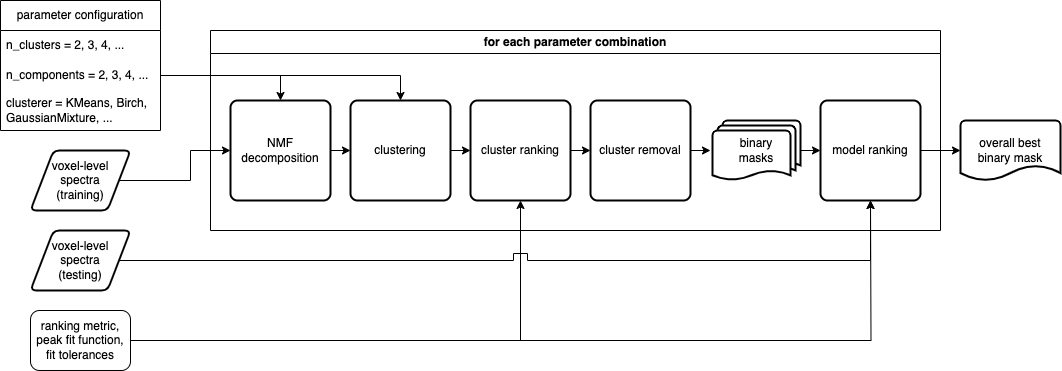}
    \caption{Overview of the NMF+clustering pipeline.}
    \label{fig:ml-pipeline-overview}
\end{figure*}

Fig.~\ref{fig:ml-pipeline-overview} gives an overview of the ML-based clustering pipeline.
First, non-negative matrix factorization (NMF) is used to decompose the voxel-level training spectra $\mathbf{X}^{[\nvox,\, \nbins]}  \geq 0$ into a lower-rank approximation with $\ncomp$ components,
\begin{align}
    \mathbf{X} \simeq \mathbf{W} \mathbf{H}
\end{align}
where $\mathbf{W}^{[\nvox,\, \ncomp]}  \geq 0$ is the matrix of weights and $\mathbf{H}^{[\ncomp,\, \nbins]} \geq 0$ is the matrix of components or feature vectors.
A regularizer of strength $\alpha_W \geq 0$ can be applied to promote sparsity in $\mathbf{W}$.
For performance, the {\tt spectre-ml} software caches the NMF decomposition for a fixed $(\ncomp, \alpha_W)$ and variable $\nclus$ rather than repeating the expensive calculation.
The NMF weights in each voxel are then used as inputs for the clustering step.
We note a number of potential advantages of clustering on NMF weights instead of some other spectral characteristic such as peak width.
NMF automatically learns low-dimensional latent structure that globally describes all available data, rather than reducing the spectral data to a single hand-chosen metric.
It is also more robust to voxels with low data, whereas extracting a width from a low-statistics peak fit may be unreliable.
Finally, NMF can be faster to compute than tens of thousands of individual peak fits.

We consider several of the standard clustering algorithms available in {\tt scikit-learn}~\cite{pedregosa2011scikit} that can scale to large numbers of samples, namely agglomerative clustering~\cite{vijaya2017review}, BIRCH~\cite{zhang1996birch}, and Gaussian mixture clustering~\cite{bouveyron2007high}, which require $\nclus$ to be specified, as well as DBSCAN~\cite{ester1996density}, OPTICS~\cite{ankerst1999optics}, and $k$-means~\cite{hartigan1979algorithm}, which determine $\nclus$ on their own.
Once the voxel clusters are generated, the desired performance metric is computed for each cluster spectrum, and each cluster is ranked by its metric.
The cluster spectra are then re-accumulated in order of their individual metrics, best to worst, and the metrics are recomputed at each accumulation.
Each such re-accumulation is one model that is then saved and can be compared against all other models to determine the best-performing voxel mask---for example, if six voxel clusters are used, five models are generated and ranked: the best cluster only, the first and second best clusters combined, and so on, up to the first through fifth best clusters combined (since including the worst cluster would then just include the entire detector).
Finally, we note that the training and testing voxel spectra do not necessarily need to be the same---a binary voxel mask computed from one dataset could be applied to a different dataset if desired.

\subsection{Heuristic clustering}
\begin{figure}[!htbp]
    \centering
    \includegraphics[width=1.0\linewidth]{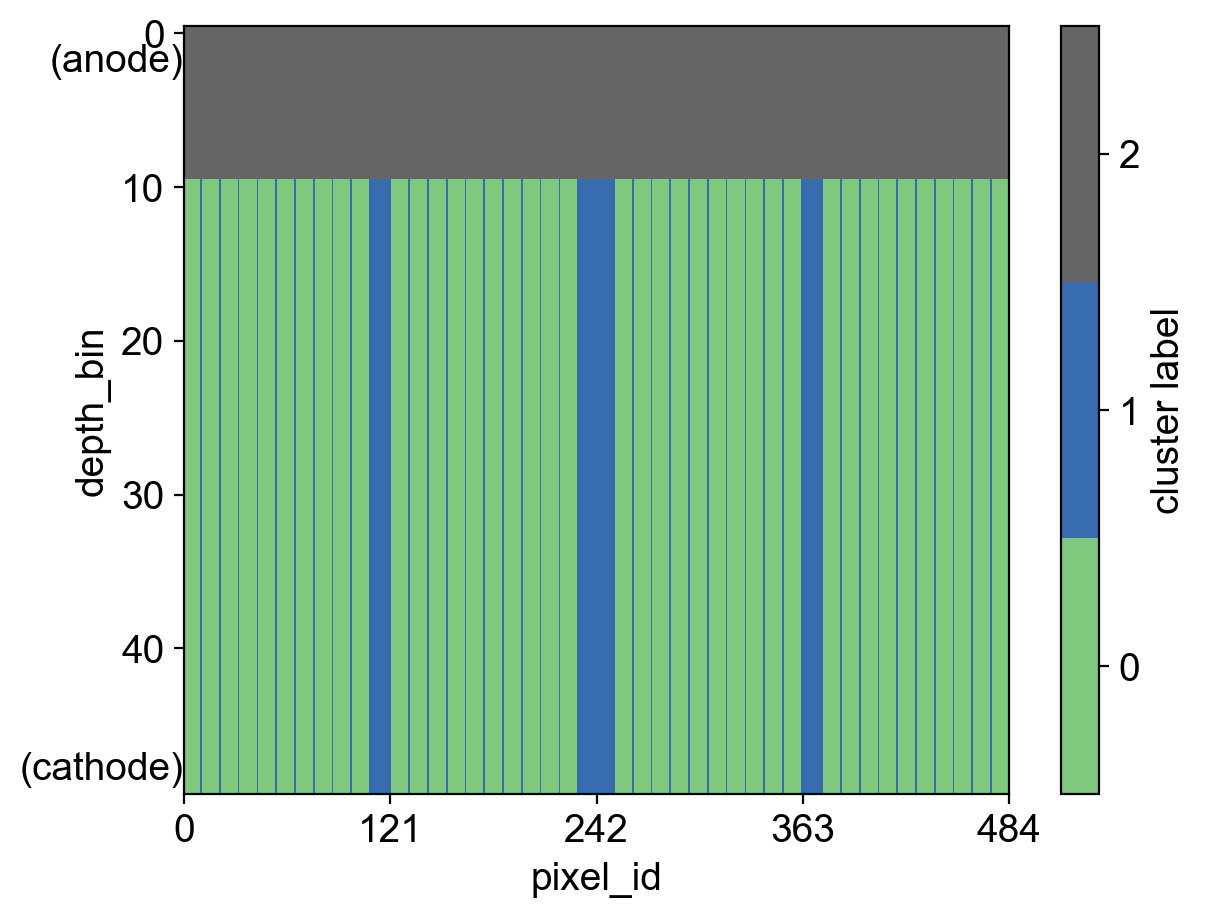}\\
    \includegraphics[width=1.0\linewidth]{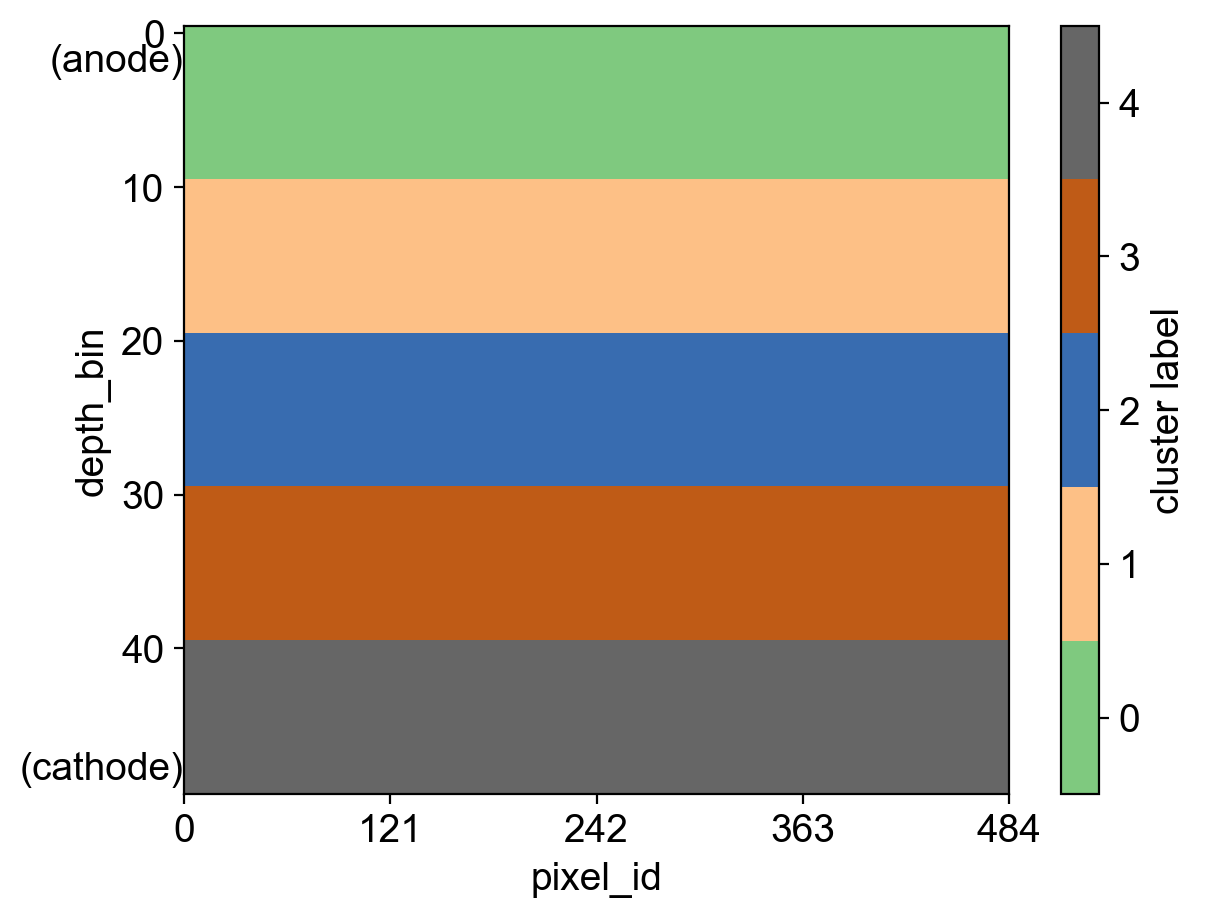}\\
    \caption{
        Example heuristic cluster assignments.
        In this and subsequent figures, the $4 \times 11 \times 11$ arrangement of pixels is unrolled into a global ``pixel\_id'' from $0$ to $483$.
        Top: edge-and-anode, with an edge width of 1 pixel and an anode depth of 10 voxels.
        Bottom: equal depth bins, with $\nclus = 5$.
    }
    \label{fig:labels_heuristic}
\end{figure}

Two additional clustering methods are based on heuristic trends in detector performance.
First, the ``edge-and-anode'' clusterer segments the detector into ``edge'', ``anode'' and ``other'' regions, and sweeps over the depth of the ``anode'' region and the width of the ``edge'' regions.
This cluster assignment scheme is based on prior characterizations of the M400 detector, where strong performance variations were found at the edges and significantly reduced performance was found near the anode.
Second, the ``equal-depth-bins'' clusterer segments the detector into $\nclus$ (approximately) equal-sized regions in depth, sweeping over $\nclus$.
Again, this scheme is a simplification of previous results in which it was observed that the ML-based clusters often form based on depth.
These two non-ML clusterers---see Fig.~\ref{fig:labels_heuristic}---are simpler than the ML-based clusterers and do not take full advantage of spatial trends in a given detector, and thus should have reduced performance.
However, they may end up being more generalizable across different M400 detectors while still retaining useful (though not optimal) performance improvements.
This transferability of models across detectors remains an ongoing study.

\subsection{Greedy ranking algorithms}
As an alternative to both the ML-based and heuristic clustering methods, we also implemented a ``greedy ranking'' algorithm, which computes the metric in every detector segment (e.g., voxel, pixel, depth bin, or detector crystal), then accumulates those spectra from best to worst, recomputing final metric values for each accumulated spectrum.
The greedy algorithm therefore ignores the fact that accumulating spectra in such a locally optimal fashion may not lead to the globally optimal result, in the expectation that the best greedy result is close to the global optimum but much cheaper to compute.
When done at the voxel level, this computation can be limited by low per-voxel statistics, since data is split across $24\,200$ voxels in the M400.
Voxels with poor statistics that cannot be sufficiently well-fit are given a metric of $+ \infty$ (assuming lower is better) and accumulated last.
Also at the voxel level, this algorithm can also be somewhat expensive, since it performs two metric calculations (in our case, two peak fits) for each voxel.
Although we perform these fits parallel via multiprocessing, voxel-level fits still typically require tens of minutes to execute.
The greedy pixel, depth bin, and detector variants are much faster and more robust to low statistics than the greedy voxel algorithm, but describe the spectra at a coarser level and thus may lose out on fine-grained information.

\subsection{Random clustering}
\begin{figure}[!htbp]
    \centering
    \includegraphics[width=1.0\linewidth]{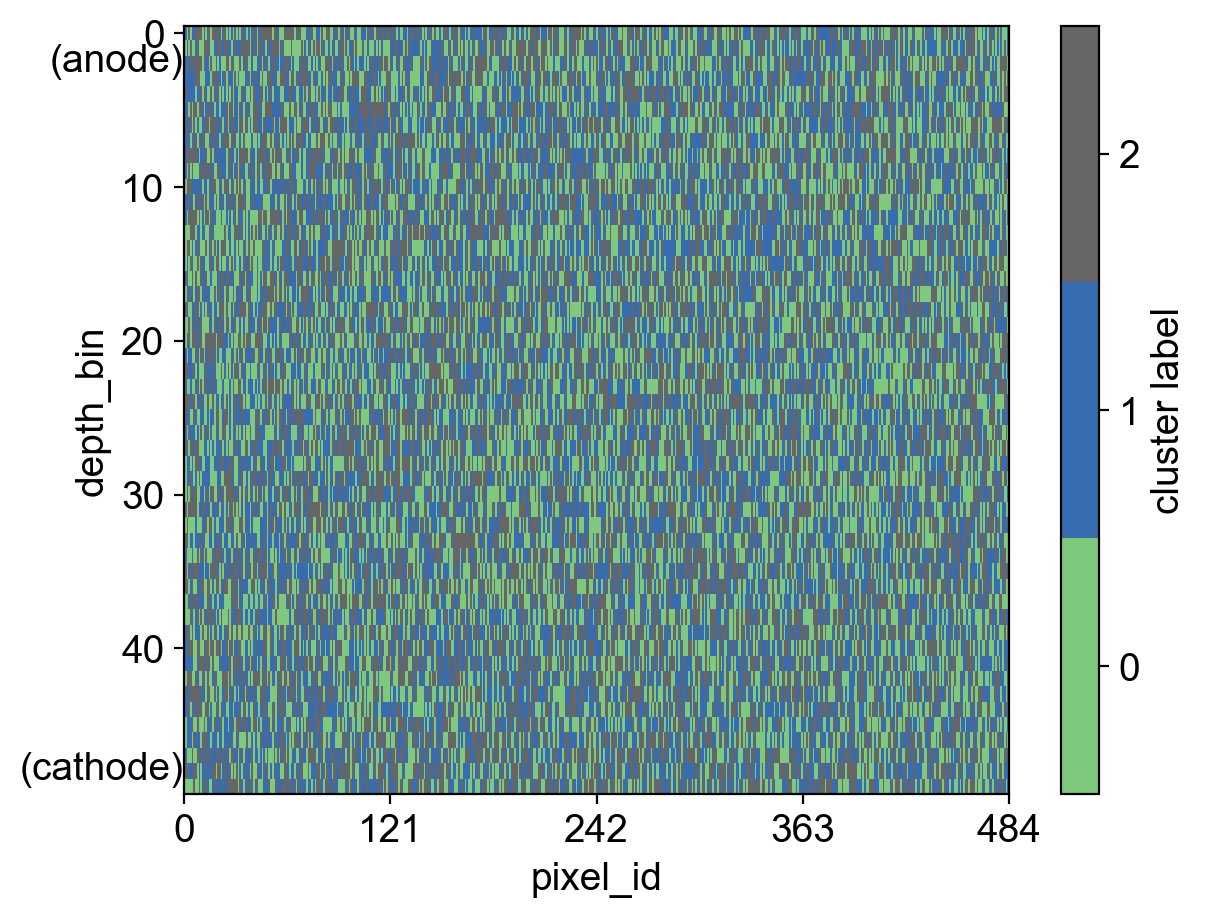}\\
    \includegraphics[width=1.0\linewidth]{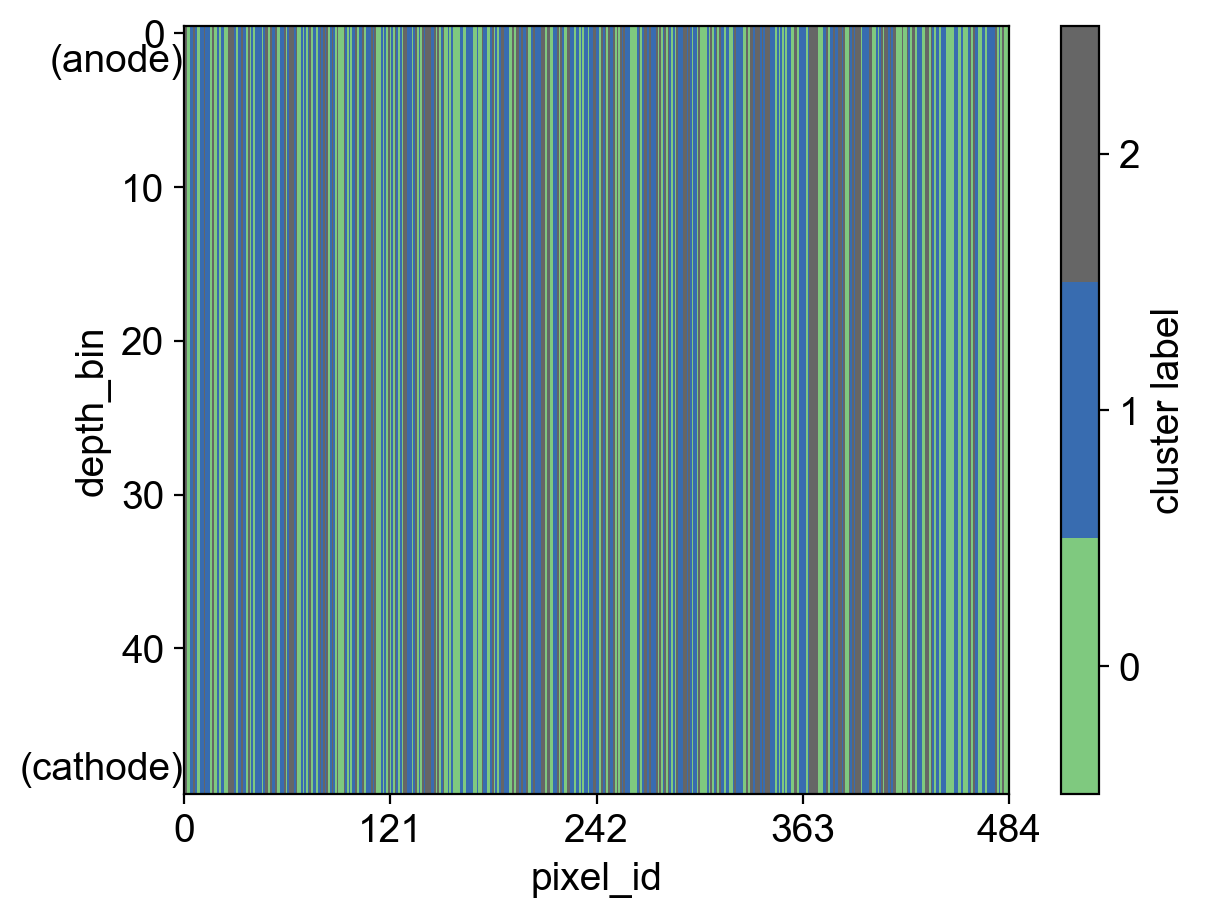}\\
    \caption{
        Example random cluster assignments.
        Top: random voxel labels, with $\nclus = 3$.
        Bottom: random pixel labels, with $\nclus = 3$.
    }
    \label{fig:labels_random}
\end{figure}

Finally, for reference, we also implement random cluster assignments, where we sweep over $\nclus$ (and the random seed) and uniformly randomly assign each voxel (or pixel or depth bin) a cluster label---see Fig.~\ref{fig:labels_random} for examples of the former two.
While these random cluster assignments are not expected to reliably generate high-performing voxel masks, they provide another useful baseline for comparison.

\subsection{Performance metrics}
The overall data-driven optimization framework is agnostic to the exact performance metric used, enabling the user to supply their own metric.
For concreteness, in this paper we consider two metrics: the relative uncertainty in a spectral peak fit parameter (typically the amplitude), and the \textit{resolvability}, a metric used to quantify the separation between closely-spaced peaks.

As we will discuss below, while the framework can support arbitrary metrics, the exact definition of the performance metric is important, and can drive the performance optimization in unexpected ways.
For instance, for metrics based on peak fits, the algorithm will not only give low ranks to overly broad spectral peaks, but also to voxel spectra that are not well-fit by the model due to, e.g., calibration shifts or secondary peak contamination.
This model fit preference can end up rejecting many voxel spectra that might intuitively be seen as ``good'' in order to optimize the fit metric.
This phenomenon is known as \textit{reward gaming} or \textit{specification gaming} and is common in artificial intelligence and machine learning optimization tasks~\cite{skalse2022defining}.

\subsubsection{Peak fit parameter relative uncertainty}
Spectral peaks in CZT detectors are non-Gaussian and asymmetric, and can be described by a Doniach peak model~\cite{doniach1970many, hyperspy_doniach} plus a background term (typically linear):
\begin{multline}\label{eq:doniach}
    f(E; \mu, A, \sigma, \gamma, c_0, c_1) =\\ c_0 + c_1 E + \frac{A \cos\left[ \pi \gamma/2 + (1-\gamma) \tan^{-1}(\epsilon / \sigma) \right]}{\left( \sigma^2 + \epsilon^2 \right)^{(1-\gamma)/2}}
\end{multline}
where
\begin{align}
     \epsilon \equiv E - \mu + \sigma \cot\left( \frac{\pi}{2-\gamma}\right),
\end{align}
$E$~is the measured energy deposition, $\mu$~is the peak centroid (i.e., the location of the peak maximum), $A$~is the peak amplitude, $\sigma > 0$ is the peak width term, $\gamma \in [0, 1)$ is the tailing or asymmetry term, and $c_0$ and $c_1$ are the intercept and slope of the background.
We note that at $\gamma = 0$, the Doniach peak shape reduces to a Lorentzian peak shape (not a Gaussian) with scale parameter~$\sigma$, and that the full-width at half-maximum of this Lorentzian is $2\sigma$.
Doniach fits are performed using the {\tt becquerel} toolkit~\cite{bandstra2021becquerel} with the {\tt lmfit}~\cite{newville2024lmfit} back-end for parameter uncertainty estimation.
We note that unlike for a Gaussian peak shape, there is no simple expression relating the Doniach amplitude parameter~$A$ to the net counts above background---in fact the integral of the Doniach part of Eq.~\ref{eq:doniach} is infinite for $\gamma > 0$~\cite{casaxps_manual, evans1991curve}.
For the purposes of this paper, however, the Doniach amplitude relative uncertainty is a useful demonstration metric, and in future work we will replace the Doniach fit with net area calculations from advanced safeguards spectral analysis codes such as GEM~\cite{berlizov2022gem, zalavadia2024uranium}.
We also assume it is representative of the dominant uncertainty in downstream NDA calculations, rather than, e.g., the efficiency uncertainty, though in principle the spatial dependence of the efficiency (and its uncertainty) could also be factored into the optimization if needed.

\subsubsection{Resolvability}
A metric was developed to quantify how well a given spectrum can resolve two closely spaced lines.
Its derivation (given in Appendix~\ref{sec:appendix}) considers two Gaussian lines of similar strengths with standard deviations equal to $\sigma$ and means separated by $\Delta \mu$.
We defined the ``resolvability'' as the signal-to-noise ratio (SNR) of a maximum likelihood estimator for the fractional difference in strength between the two lines---essentially, how well can the two lines be spectroscopically quantified for purposes of assay given their extent of overlap.
Fisher information was used to estimate the variance of the hypothetical estimator.
Neglecting the presence of background, and expanding to lowest order in $\Delta \mu / \sigma$, the variance was found to be proportional to $(\Delta \mu / \sigma)^{-2} / A$, where~$A$ is the line strength.
Factoring out terms that are constant and keeping only terms that are measurable spectral properties, the resolvability metric is proportional to the inverse of the square root of the variance, i.e.,
\begin{align}\label{eq:resolvability}
    r &\equiv A^{1/2} / \sigma.
\end{align}
In this work, instead of maximizing the resolvability, we minimize its inverse.
The resolvability is an intuitive spectral performance metric as it penalizes the peak width~$\sigma$ and prefers the peak amplitude~$A$; moreover, it captures the $\sqrt{N}$ improvement trend expected from Poisson statistics.
Although it was initially derived for separating two closely-spaced peaks, its intuitive nature and similarity to a signal-to-background ratio means it can also be useful for optimizing an isolated peak on top of background, as shown later in Section~\ref{sec:example_alpha}.
By no coincidence, the resolvability is extremely closely related to Lehr's rule of thumb for the t-test~\cite{lehr1992sixteen}, which approximates the minimum sample size~$n$ to achieve a statistical power of $80\%$ at a significance level of $0.05$ as
\begin{align}
    n \geq 16 s^2 / d^2,
\end{align}
where $s^2$ is an estimate of the population variance and $d^2$ is the squared difference in means to be detected.
Holding~$d^2$ constant, the power of the test can be improved by increasing $n/s^2$, which is proportional to the square of the resolvability in Eq.~\ref{eq:resolvability}.

\section{Results}\label{sec:results}
In this section, we present four example optimizations covering various spectral performance metrics, source spectra, and trends in analysis.
The examples range in complexity from an isolated photopeak from a long-dwell Eu-154 calibration source measurement using the inverse resolvability metric to a short-dwell Pu doublet peak with a performance metric based on the amplitude peak fit relative uncertainty.
The examples also comprise three different M400 units---one from Lawrence Berkeley National Laboratory (LBNL), one from Los Alamos National Laboratory (LANL), and a loaner detector from the vendor.
Results are summarized in Table~\ref{tab:results-summary}, and explored in more detail in the following sub-sections.

\begin{table*}[!htbp]
    \centering
    \caption{Summary of optimization results for each example}
    \begin{tabular}{c||c|c|c|c}
Feature & Example~1a & Example~1b & Example~2 & Example~3 \\\hline\hline
photopeak & 123 keV Eu-154 & 1274 keV Eu-154 & 186 keV U & 204 keV Pu \\
detector & LBNL & LBNL & loaner & LANL \\
dwell time & 64 hours & 64 hours & 49 min & 400 min \\\hline
metric & inv.\ resolv.\ & inv.\ resolv.\ & peak amp.\ rel.\ unc.\ & peak amp.\ rel.\ unc.\ \\
clusterers & all 6 & all 6 & Gaussian, Agglomerative & all 6 \\
$\ncomp$ & 1--7 & 1--7 & 1--6 & 1--6 \\
$\nclus$ & 2--7 & 2--7 & 2--6 & 2--7 \\
\# parameter combs & 2481 & 2481 & 1814 & 2403 \\
\# unique models & 30584 & 29362 & 27244 & 29444 \\
runtime$^*$ & 2.5 hours & 2.5 hours & 1.5 hours & 2.5 hours \\\hline
bulk metric & $8.320 \times 10^{-5}$ & $3.585 \times 10^{-3}$ & $2.36\%$ & $1.23\%$ \\ \hline
best ML clusterer & Agglomerative Clustering & K-Means & Gaussian Mixture & Agglomerative Clustering \\
best ML metric & $7.499 \times 10^{-5}$ & $3.253 \times 10^{-3}$ & $0.83\%$ & $1.01\%$ \\
rel eff at best ML metric & $0.9048$ & $0.8522$ & $0.2420$ & $0.6286$ \\ \hline
best greedy algorithm & depth\_bin & pixel & pixel & voxel \\
best greedy metric & $7.503 \times 10^{-5}$ & $3.524 \times 10^{-3}$ & $1.06\%$ & $0.87\%$ \\
rel eff at best greedy metric & $0.9044$ & $0.9258$ & $0.1658$ & $0.2780$ \\
    \end{tabular}
    \vspace{4pt}
    \\$^*$includes ${\sim}30$~minutes for the greedy voxel algorithm
    \label{tab:results-summary}
\end{table*}

While more detailed runtimes are given each sub-section, the ML pipeline typically takes $2$--$3$~hours on a 2019 MacBook Pro with a $2.4$~GHz 8-Core Intel Core i9 processor and $64$~GB of RAM, depending on the breadth of the parameter sweep configured by the user.
Additional walltime improvements could be realized by distributing parameter combinations in parallel over a compute cluster, though we note that some of the underlying {\tt scikit-learn} algorithms already distribute numerical work over cores.

\subsection{Examples 1a and 1b---Eu-154 calibration source}\label{sec:example_alpha}

Examples~1a and 1b consist of optimizing the inverse resolvability metrics of the $123$~keV and $1274$~keV photopeaks, respectively, in a $64$-hour-long Eu-154 calibration source measurement with the LBNL M400 detector.
The parameter sweep considered $\nclus = 2$--$7$, $\ncomp = 1$--$7$, $\alpha_W \in [0, 0.01, 0.10]$, all six ML-based clustering methods, the equal-depth-bin clusterer for each $\nclus$, the edge-and-anode clusterer with $\nanode = 0$--$24$ and $\nedge=0$--$4$, $100$ random pixel, voxel, and depth bin combinations, and the four greedy algorithm variants.
This resulted in $2481$~parameter combinations and a total of ${\sim}30$k~unique models tested, which ran in ${\sim}2.5$~hours on the aforementioned hardware.

\begin{figure*}[!htbp]
    \centering
    \includegraphics[width=0.40\linewidth]{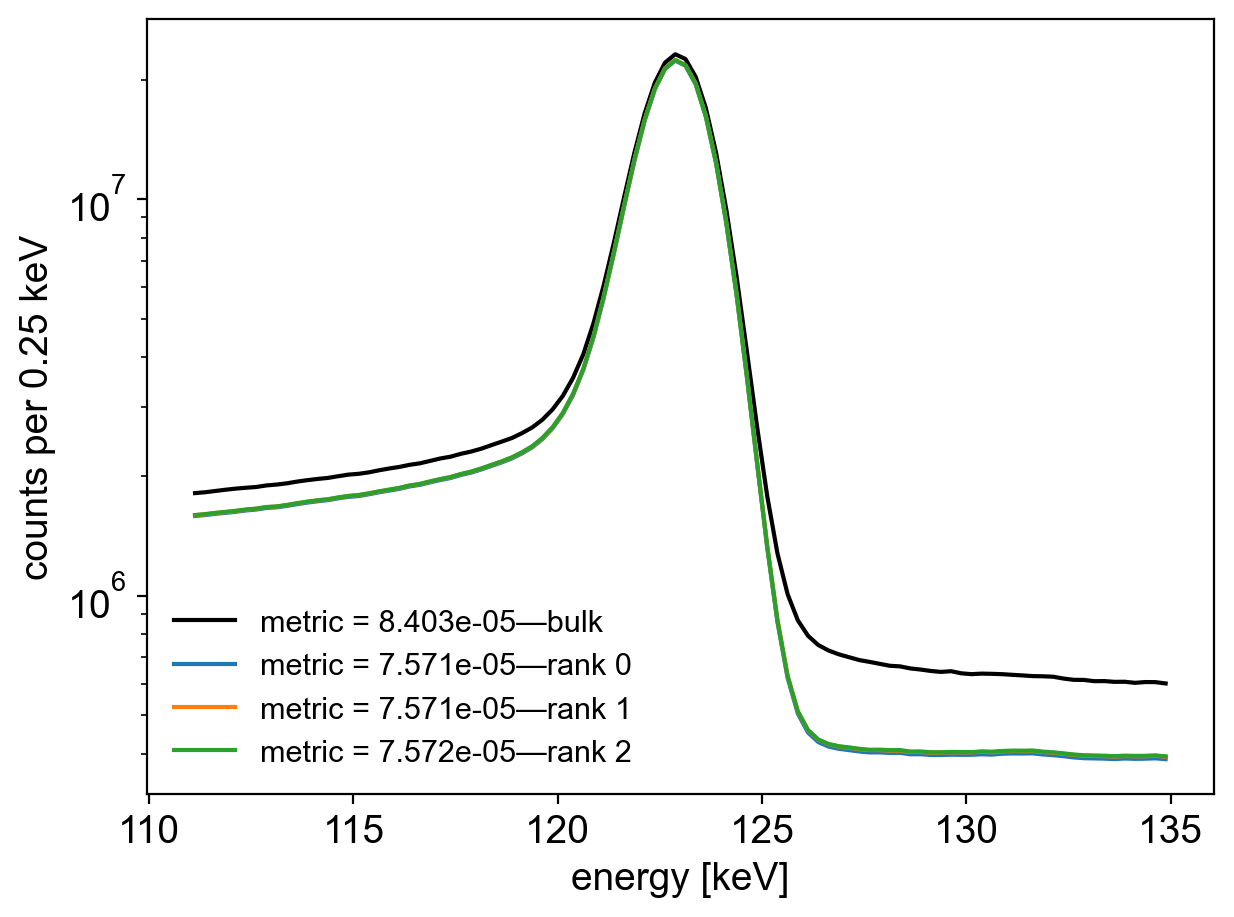}
    \includegraphics[width=0.40\linewidth]{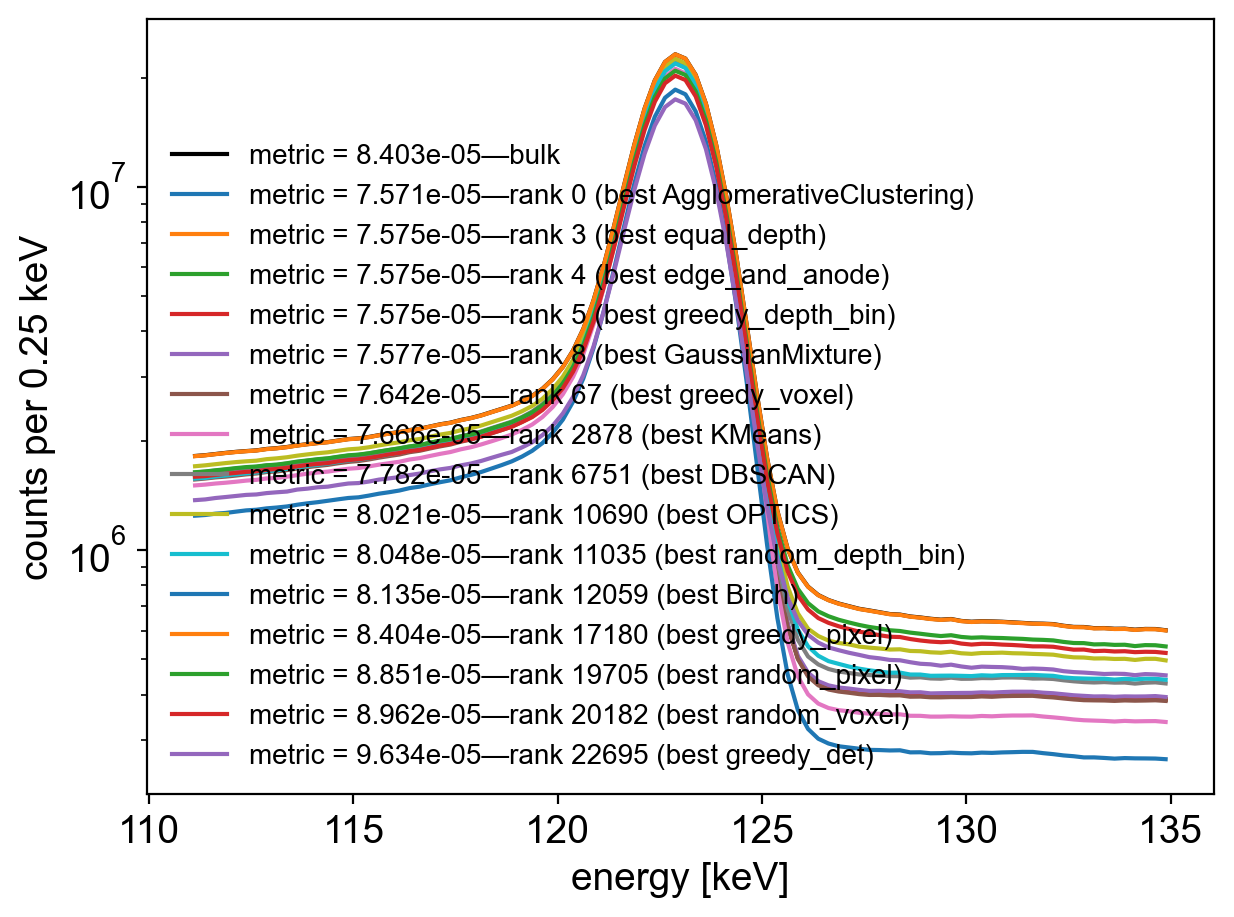}\\
    \includegraphics[width=0.40\linewidth]{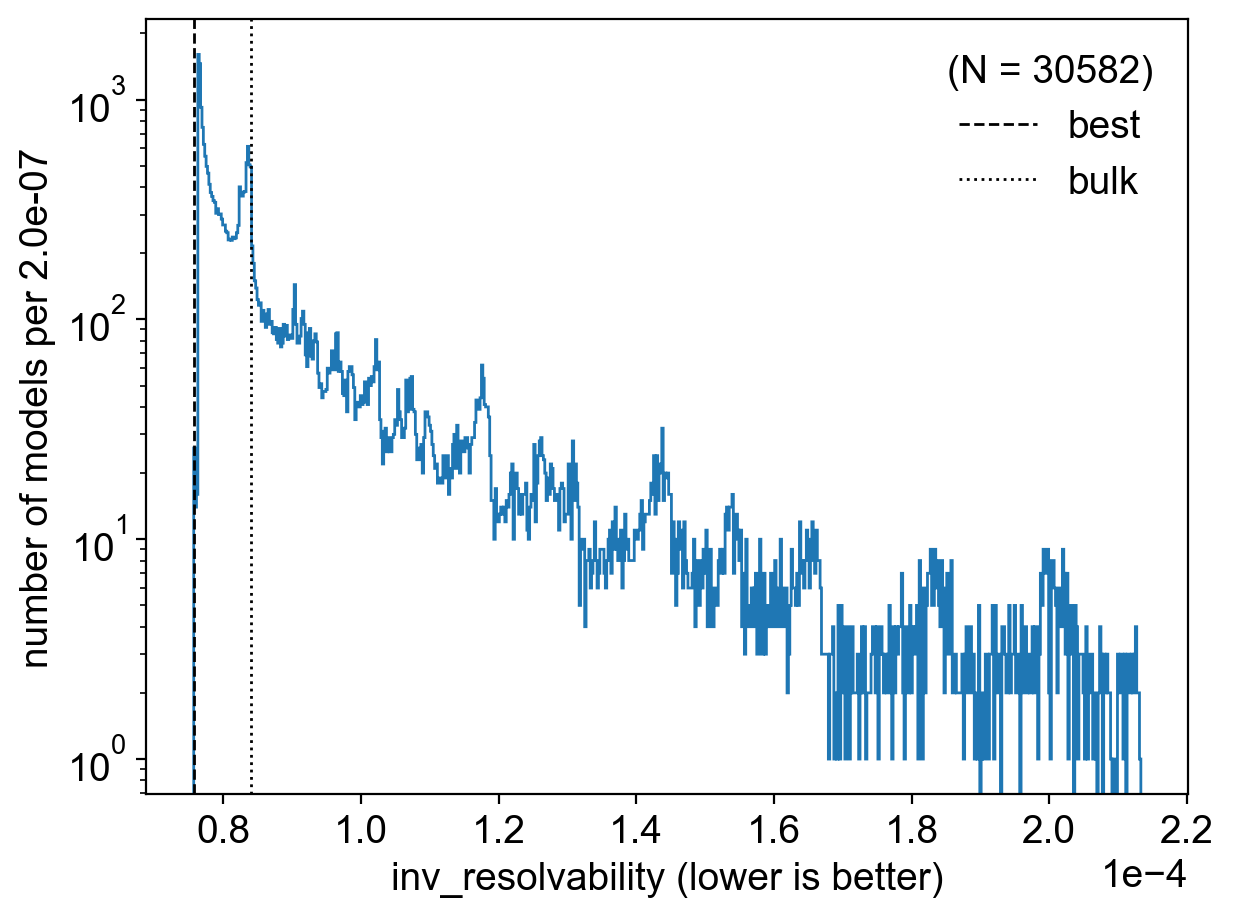}
    \includegraphics[width=0.40\linewidth]{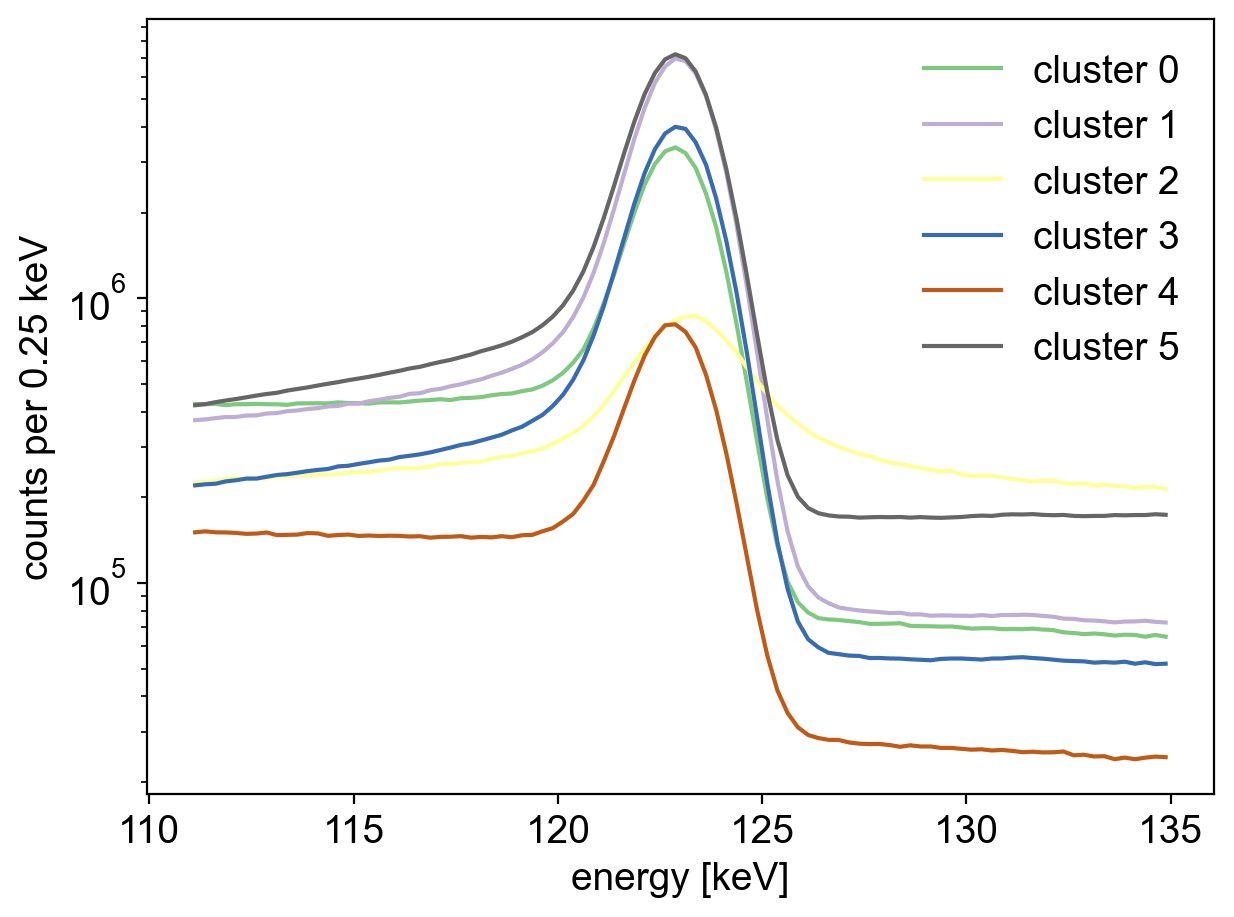}\\
    \includegraphics[width=0.40\linewidth]{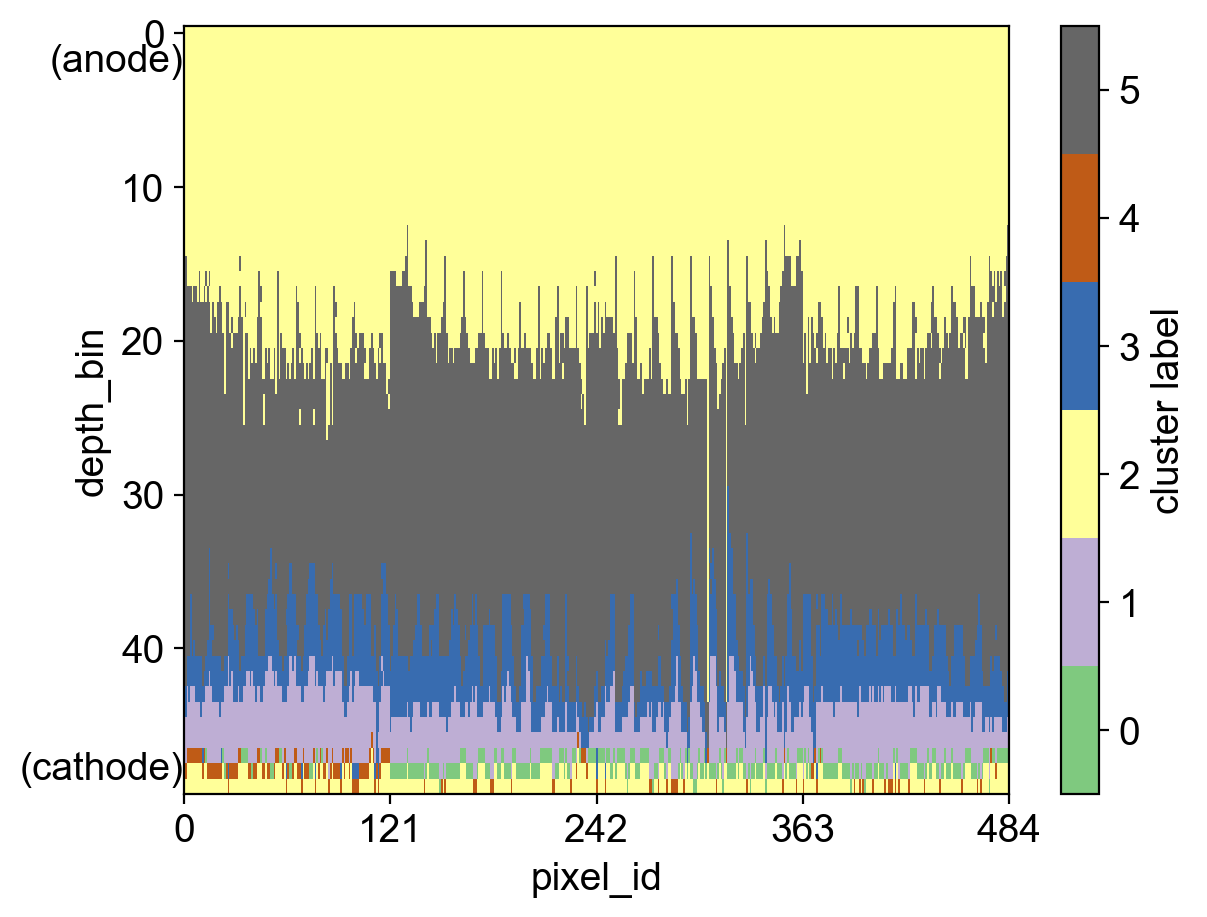}
    \includegraphics[width=0.40\linewidth]{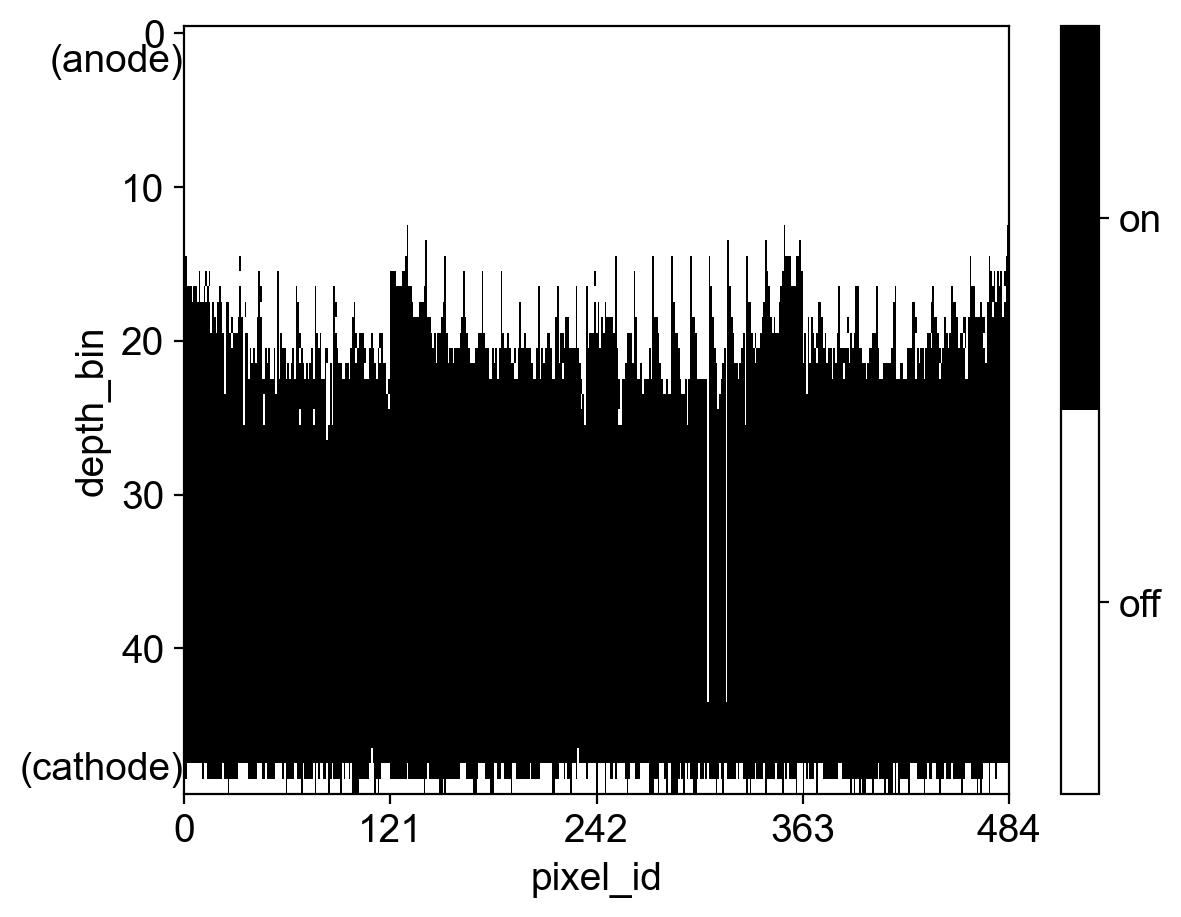}\\
    \includegraphics[width=0.80\linewidth]{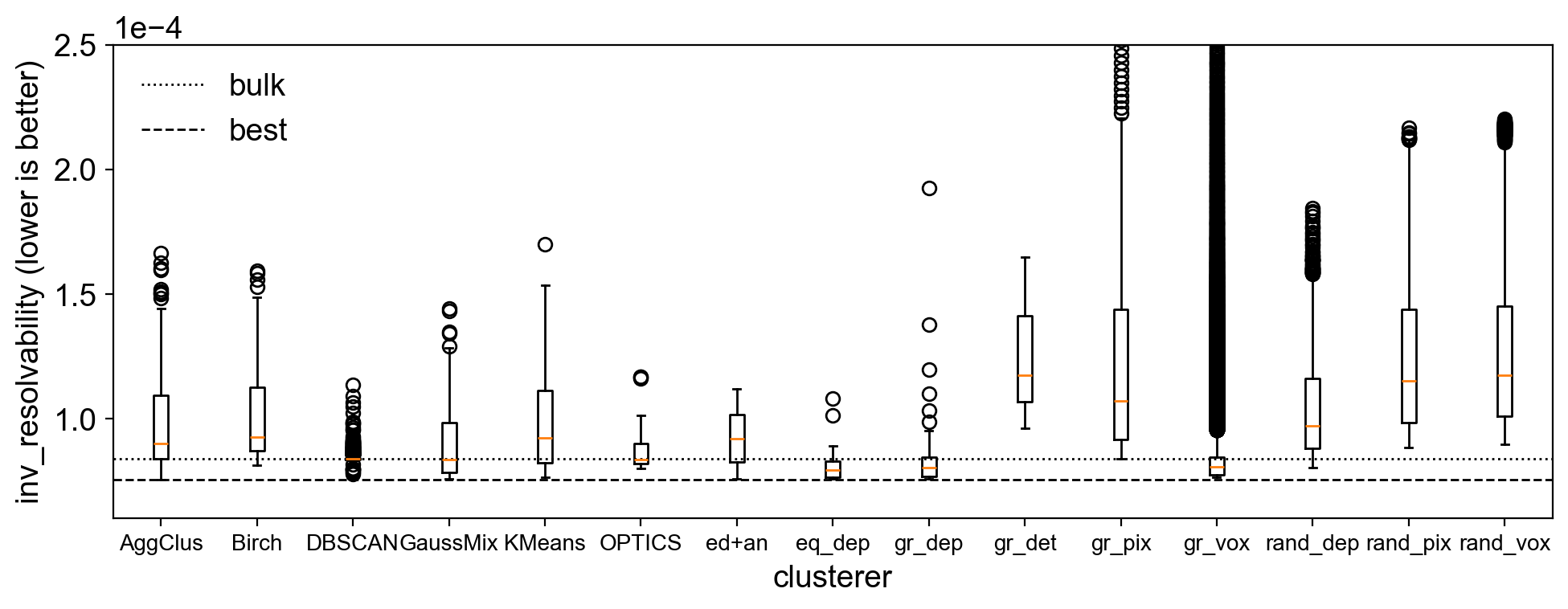}\\
    \caption{
        Optimization results for the $123$~keV Eu-154 peak in Example~1a.
        Top left: the three best-inverse-resolvability spectra compared to the bulk (unoptimized) spectrum.
        The blue, orange, and green curves all overlap.
        Top right: the top spectrum from each clustering method.
        Upper left: histogram of metric values from all tested models.
        Upper right: spectrum in each cluster in the optimized result.
        Lower left: cluster labels in the optimized result.
        Lower right: voxel mask in the optimized result.
        Bottom: distribution of metric values for each clusterer.
    }
    \label{fig:example_alpha}
\end{figure*}

In the $123$~keV case, Fig.~\ref{fig:example_alpha} shows that the bulk inverse resolvability of $8.320 \times 10^{-5}$ is improved to $7.571 \times 10^{-5}$ when using $\ncomp = 4$, $\alpha_W = 0.01$, $\nclus = 6$ via AgglomerativeClustering, and removing only one cluster ($\# 2$).
This $10\%$ relative improvement comes at the cost of reducing the detector relative efficiency to $90\%$, and appears to manifest as a noticeable reduction in the low- and especially high-energy background on either side of the peak.
For simplicity, the relative efficiency in this and subsequent analyses is estimated as the ratio of total counts within the input spectrum energy range relative to that of the bulk detector, rather than using any dedicated peak fits and/or background subtraction.
We also note that the clusters found here are highly correlated with depth bin.
While most models tested produce results similar to or worse than bulk unoptimized spectrum, there is a small tail of results similar to the best (AgglomerativeClustering) result, including the best edge-and-anode, equal-depth, and GaussianMixture results.
The Birch, DBSCAN, OPTICS, and both random clusterers tend to perform worse than the Agglomerative, Gaussian Mixture, edge-and-anode, and equal-depth-bin clusterers.
The top three models all use AgglomerativeClustering to remove only one cluster but differ slightly in terms of their $(\nclus, \ncomp, W_\alpha)$---from best to worst, $(6, 4, 0.01)$, $(6, 4, 0.10)$, and $(5, 7, 0.00)$.

\begin{figure}[!htbp]
    \centering
    \includegraphics[width=1.0\linewidth]{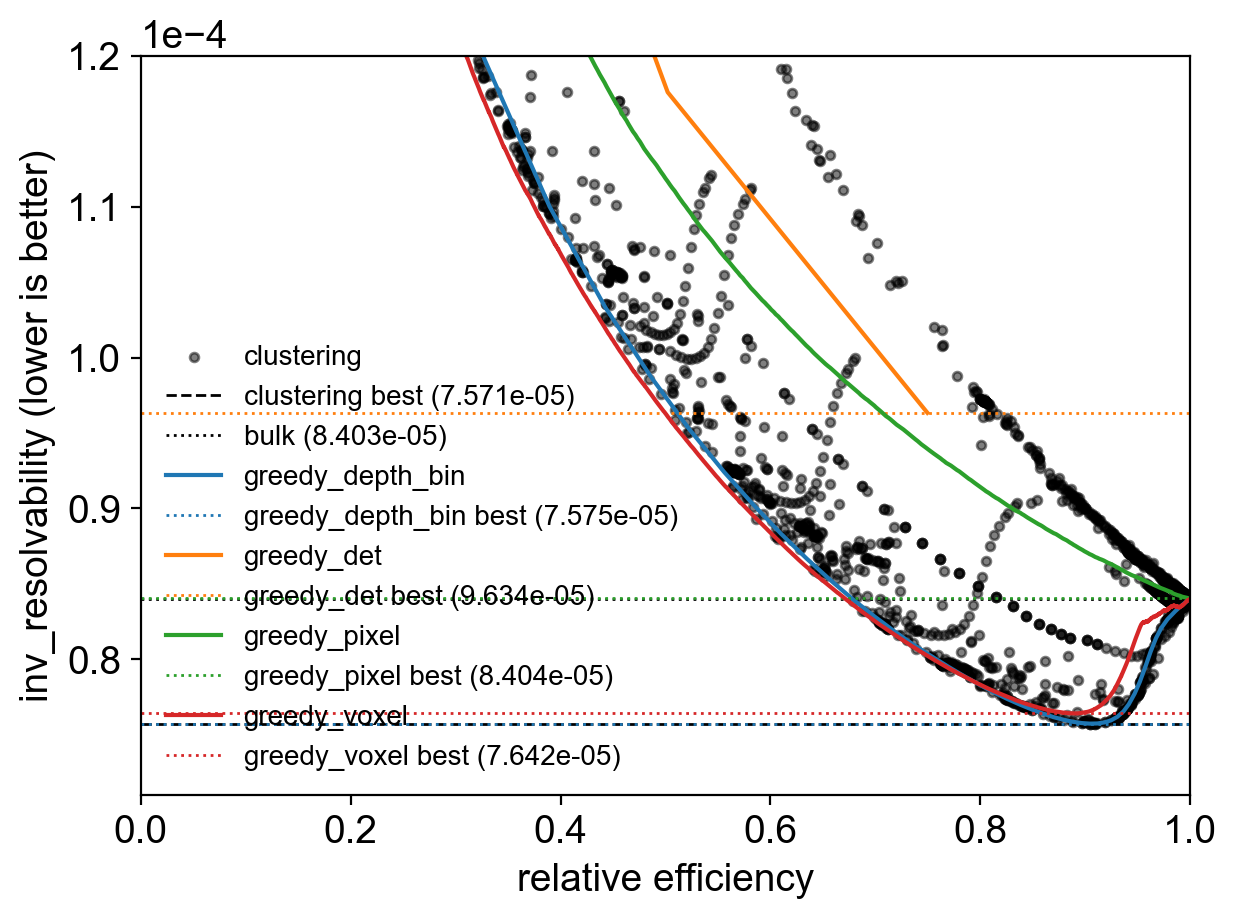}\\
    \includegraphics[width=1.0\linewidth]{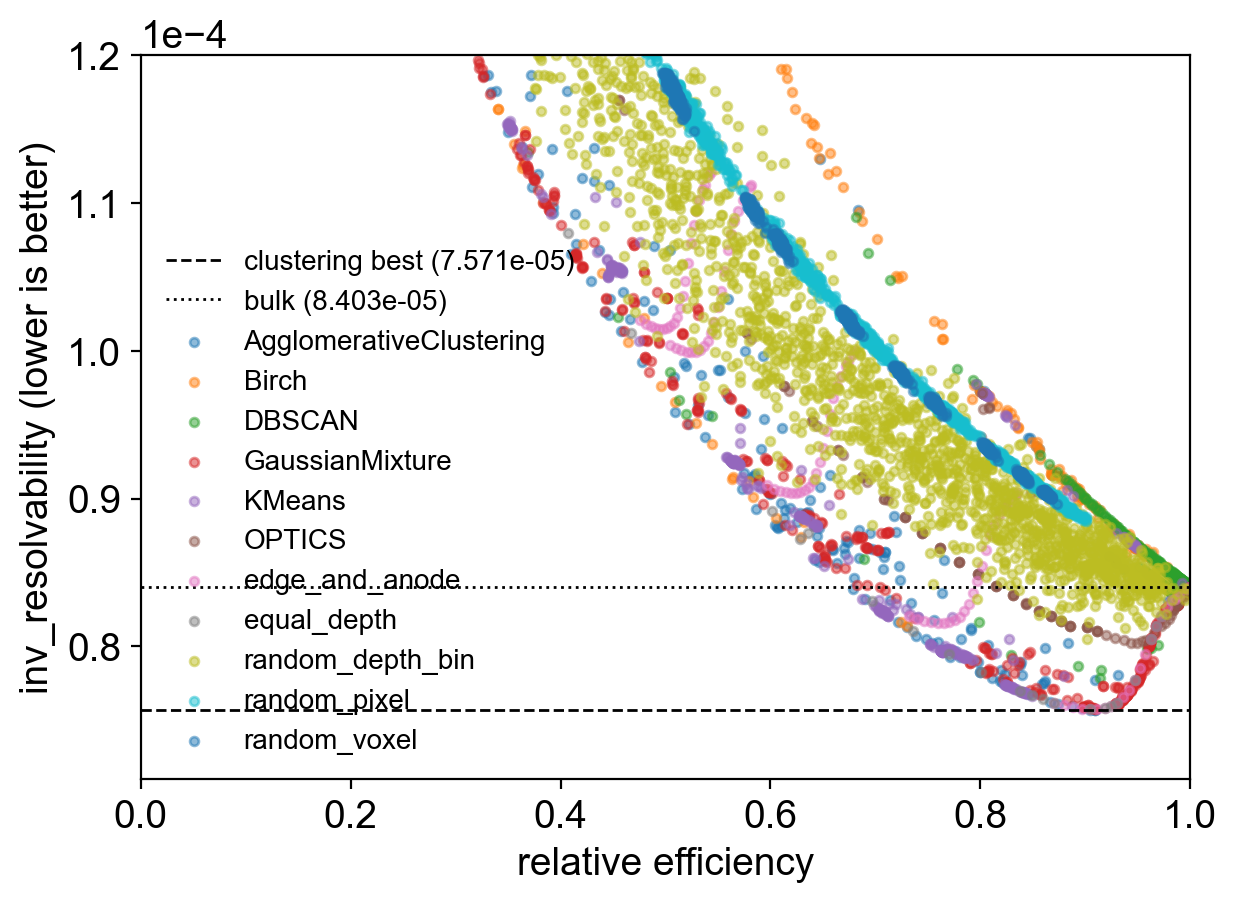}
    \caption{
        Inverse resolvability vs.\ relative efficiency for Example~1.
        The axes limits have been zoomed to focus on lower (better) inverse resolvabilities.
        Top: all SPECTRE-ML models (black dots) and greedy models (solid lines).
        Points from the random clusterers have been excluded for clarity.
        Bottom: SPECTRE-ML models colorized by clusterer type.
    }
    \label{fig:example_alpha_metric_vs_rel_eff}
\end{figure}

Furthermore, Fig.~\ref{fig:example_alpha_metric_vs_rel_eff} shows the inverse resolvability from Example~1a vs.\ relative detector efficiency.
In general, since resolvability is proportional to efficiency, the inverse resolvability vs.\ efficiency tends to follow a ${\sim}1/x$ shape.
However there is clear structure within the plot that is evident when the data is further broken down by clusterer type.
Fig.~\ref{fig:example_alpha_metric_vs_rel_eff} also compares results against the four greedy algorithm variants.
The greedy pixel and detector algorithms never improve upon the bulk, let alone the SPECTRE-ML model(s) at similar efficiency.
The greedy voxel algorithm initially performs better than the greedy depth bin, but the trend reverses at an efficiency of ${\sim}0.9$.
While the greedy voxel and greedy depth bin algorithms often slightly outperform SPECTRE-ML at lower efficiencies, SPECTRE-ML indeed attains a slightly better final metric value.
The greedy voxel algorithm ran in ${\sim}30$~min while the greedy depth bin version took only $3$~seconds.

\begin{figure*}[!htbp]
    \includegraphics[width=0.49\textwidth]{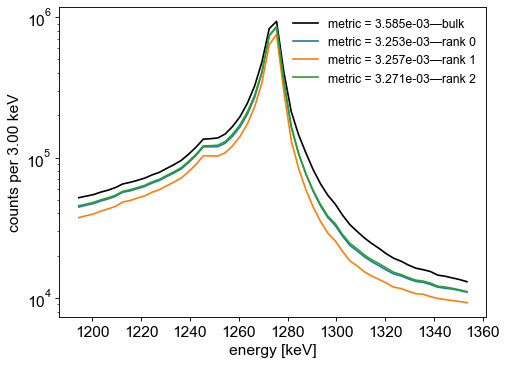}
    \includegraphics[width=0.49\textwidth]{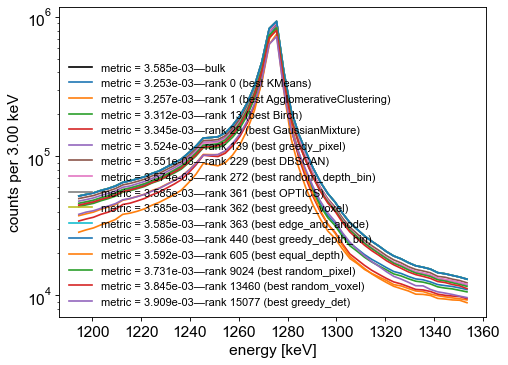}\\
    \includegraphics[width=0.49\textwidth]{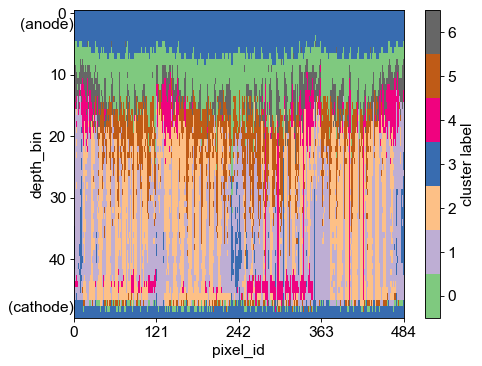}
    \includegraphics[width=0.49\textwidth]{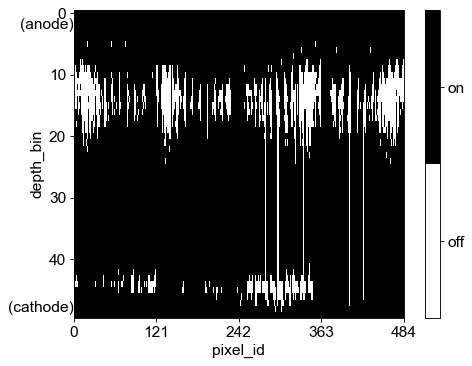}
    \caption{
        Optimization results for the $1274$~keV Eu-154 peak in Example~1b.
        Top left: the three best-inverse-resolvability spectra compared to the bulk (unoptimized) spectrum.
        Top right: the top spectrum from each clustering method.
        Bottom left: cluster labels in the optimized result.
        Bottom right: voxel mask in the optimized result.
    }
    \label{fig:example_delta}
\end{figure*}

Example~1b demonstrates the optimization at higher peak energies---in the $1274$~keV case, Fig.~\ref{fig:example_delta} shows that the bulk inverse resolvability of $3.585 \times 10^{-3}$ is improved to $3.253 \times 10^{-3}$ when using $\ncomp = 3$, $\alpha_W = 0.1$, $\nclus = 7$ via $k$-means clustering, and removing $2$ clusters ($\# 6$ and $\# 4$).
Similar to Example~1a at the $123$~keV peak of the same spectrum, a $10\%$ relative improvement is achieved by reducing the relative efficiency to $93\%$ of the bulk detector.
Although $k$-means provides the single best model, Agglomerative Clustering again performs well overall, giving the second best model and five of the top $20$.
In contrast to the $123$~keV peak, here the greedy depth bin algorithm does not improve over the bulk, and the best greedy algorithm (pixel) offers only a marginal improvement.

\subsection{Example 2---uranium sample}\label{sec:example_beta}

Example~2 minimizes the relative uncertainty of the Doniach peak amplitude fit parameter in the $185.7$~keV peak of U-235 in a $49$~minute measurement of a $93\%$-enriched U$_3$O$_8$ sample with the loaner M400 detector.
The parameter sweep here was smaller than those in Example~1, using $\nclus = 2$--$6$ and $\ncomp = 1$--$6$ and only two of the six {\tt scikit-learn} clustering algorithms (AgglomerativeClustering and GaussianMixture).
This resulted in $1814$ total parameter combinations and $27244$ total models tested, and ran in ${\sim}1.5$~hours.

\begin{figure*}[!htbp]
    \centering
    \includegraphics[width=0.49\linewidth]{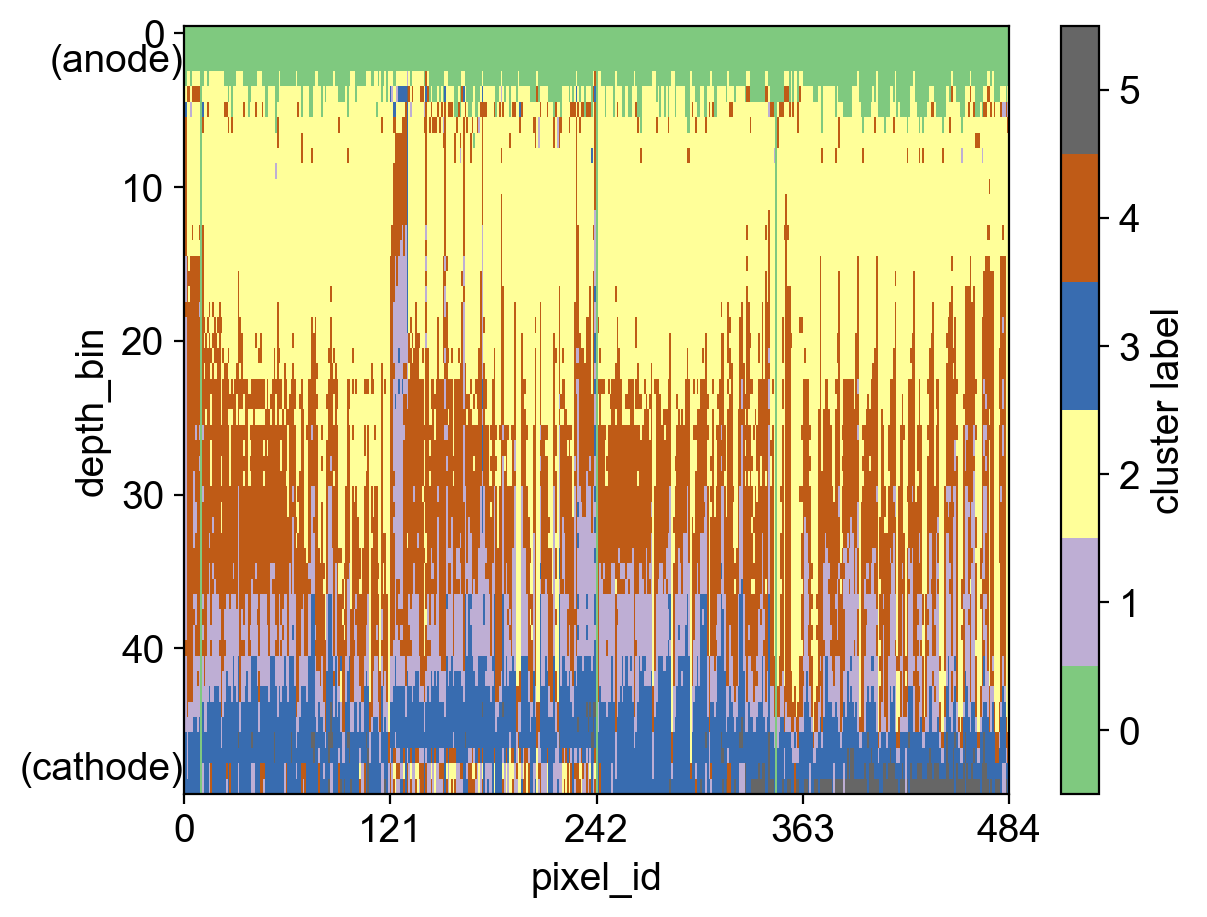}
    \includegraphics[width=0.49\linewidth]{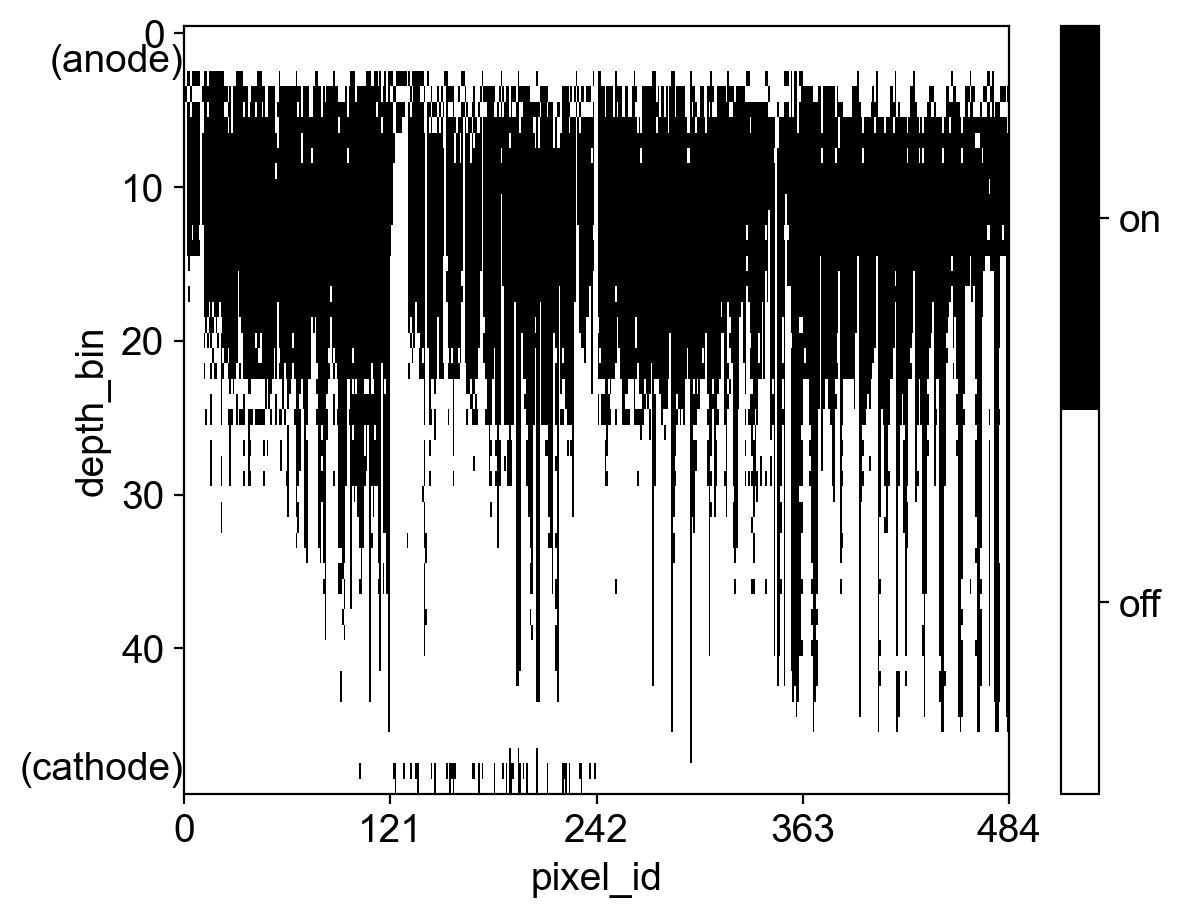}\\
    \includegraphics[width=0.49\linewidth]{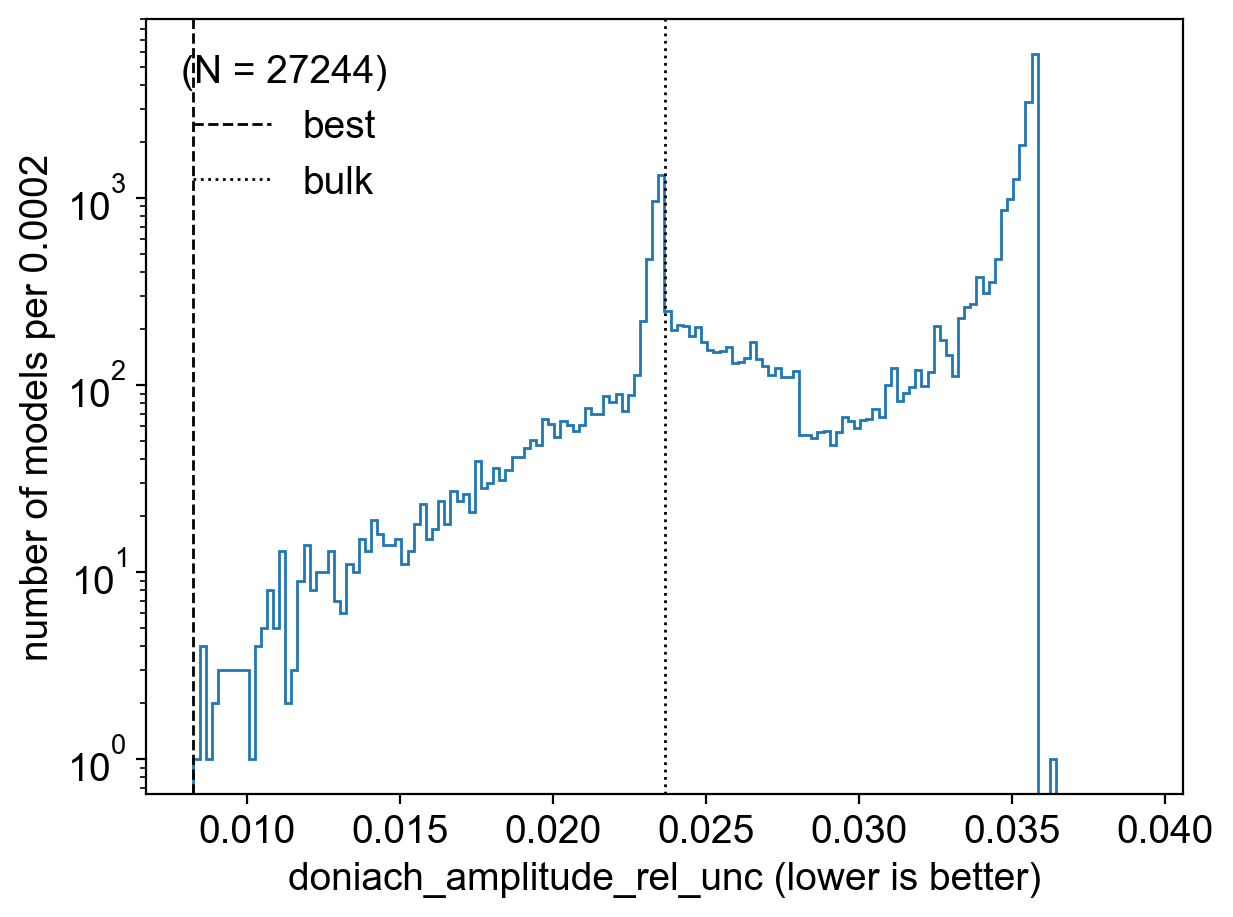}
    \includegraphics[width=0.49\linewidth]{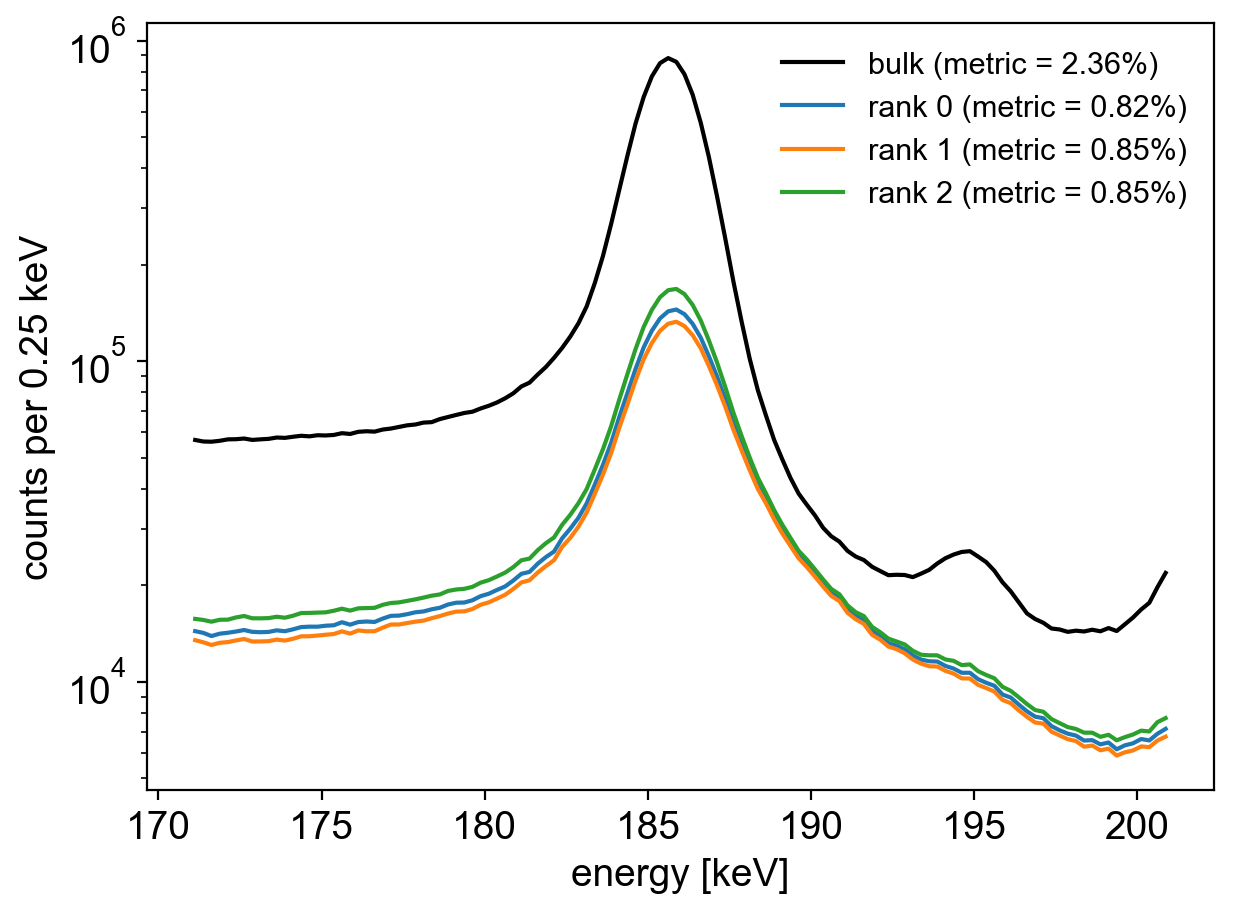}\\
    \includegraphics[width=0.49\linewidth]{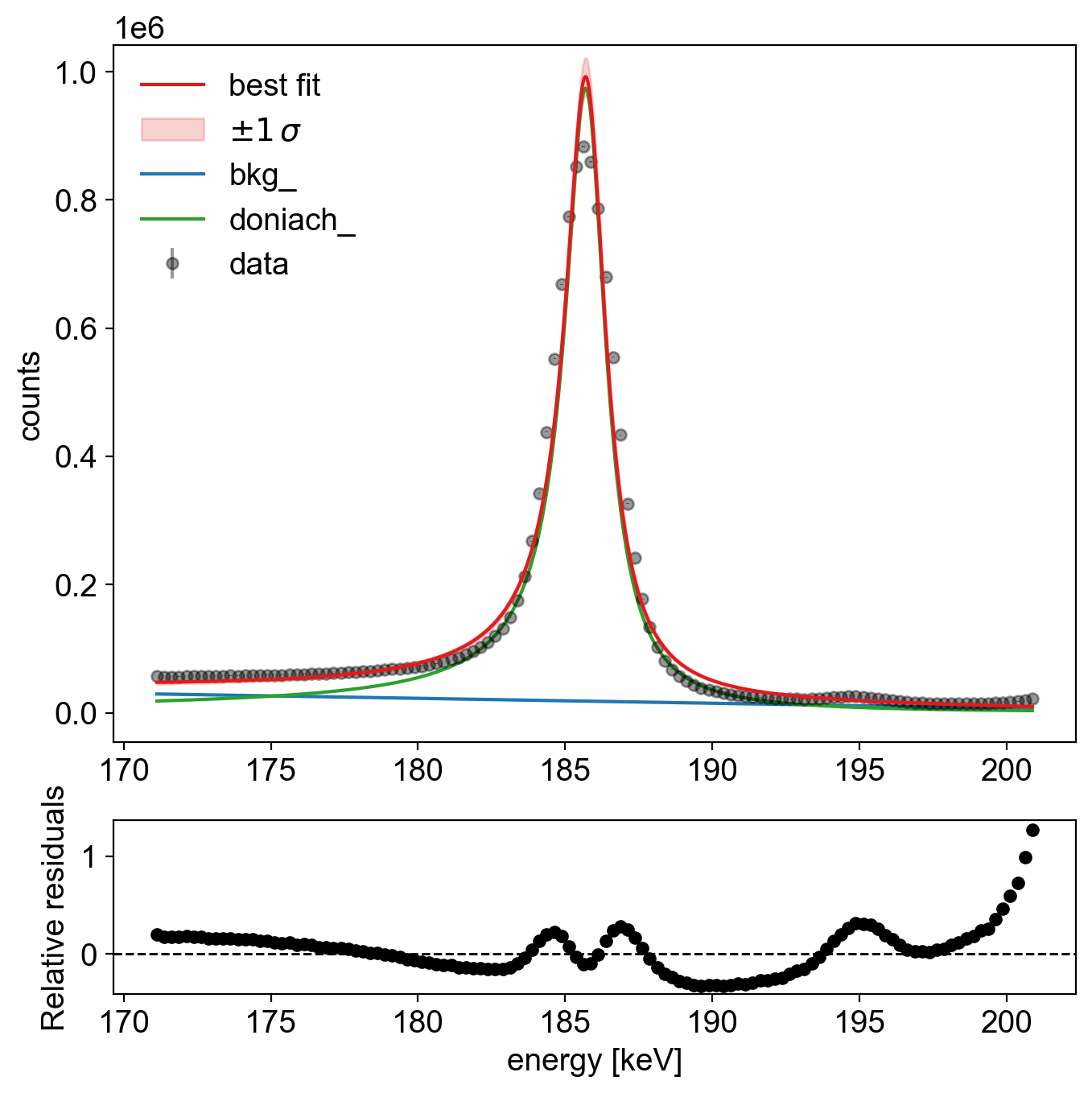}
    \includegraphics[width=0.49\linewidth]{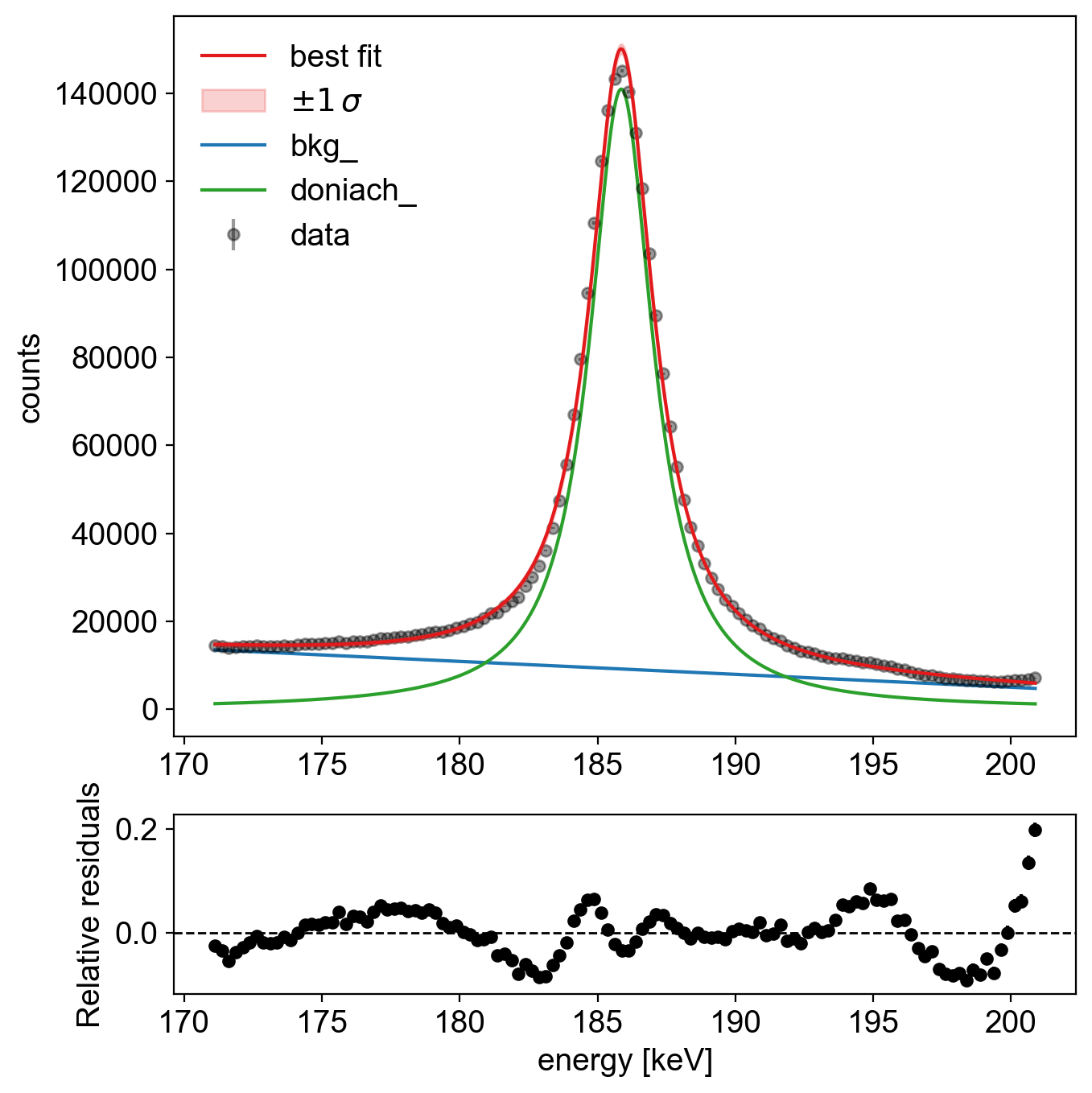}
    \caption{
        Optimization results for the $185.7$~keV U-235 peak with the loaner detector.
        Top left: best cluster labels.
        Top right: best cluster mask.
        Middle left: histogram of all metric values.
        Middle right: bulk and top~$3$ spectra.
        Bottom left: bulk peak fit.
        Bottom right: best peak fit.
    }
    \label{fig:example_beta}
\end{figure*}

Fig.~\ref{fig:example_beta} shows that the bulk Doniach amplitude relative uncertainty of $2.36\%$ is improved to $0.83\%$ when using $1/6$~Gaussian Mixture clusters with $\ncomp = 1$ and $\alpha_W = 0$.
This $2.9 \times$ relative improvement comes at the cost of a reduction in detector relative efficiency to $24\%$, and stems largely from improving the goodness-of-fit to the Doniach peak shape.
In particular, the bulk peak fit substantially overshoots the data at the peak centroid, undershoots the data just outside the centroid, and continues to have deviations in its tails.
The best peak fit by contrast fits the data much more closely throughout the energy domain, especially in the high-energy tail, where the best spectrum has a much smaller $195$~keV peak contribution, which reduces the systematic model error and improves the metric.
We also note that the best peak fit is significantly more Lorentzian, with a fit asymmetry term of $\gamma = 3 \times 10^{-10} \pm 4 \times 10^{-3}$ (consistent with zero), compared to the bulk $\gamma = 0.072 \pm 0.009$.
Finally, although we used the Doniach peak fit relative uncertainty as a convenient optimization target, we note that direct calculation of the net fit area with full correlated error propagation gives area relative uncertainties of $2.47\%$ (bulk) and $1.57\%$ (best), in rough agreement with the improvement in Doniach amplitude relative uncertainties.

\begin{figure}[!htbp]
    \centering
    \includegraphics[width=1.0\linewidth]{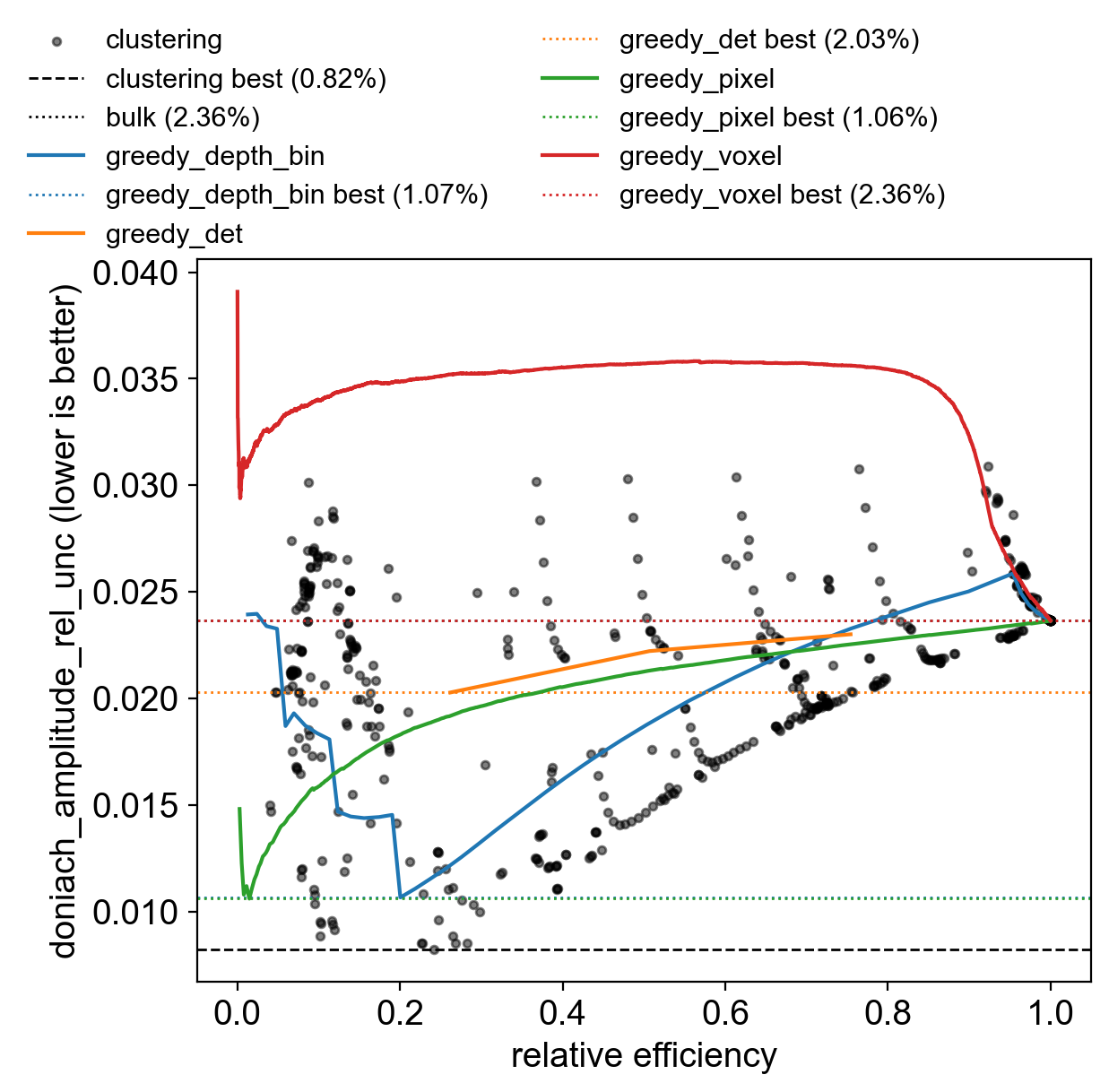}
    \caption{
        Doniach amplitude relative uncertainty vs.\ relative efficiency for Example~2.
        Points from the random clusterers have been excluded for clarity.
    }
    \label{fig:example_beta_metric_vs_rel_eff}
\end{figure}

The greedy algorithm results in this example show different trends from those in Example~1 likely due to the change in performance metric.
Fig.~\ref{fig:example_beta_metric_vs_rel_eff} shows the SPECTRE-ML and greedy metric values vs.\ relative efficiency, as in Fig.~\ref{fig:example_alpha_metric_vs_rel_eff}.
Here, however, the greedy voxel algorithm performs poorly, maintaining the highest relative uncertainty of nearly any model across the efficiency domain, while the greedy pixel and detector algorithms largely get worse with increasing data.
The greedy pixel algorithm reaches the best metric of any of the four greedy variants with a value of $1.06\%$ at a relative efficiency of only $1.5\%$.
By contrast, the greedy depth bin algorithm begins to improve with increasing data but reaches its best value of $1.07\%$ at a larger efficiency of $17\%$ before degrading towards the bulk metric value.
These trends, coupled with the SPECTRE-ML points that show degradation in uncertainty with increasing relative efficiency above ${\sim}25\%$, suggest that the uncertainty is primarily driven by fit error rather than statistical uncertainty.
The best models thus tend to remove a higher fraction of voxels than in Example~1 in order to minimize inter-crystal or inter-pixel peak shape changes that can drive up the systematic fit error.

\subsection{Example 3---plutonium sample}

Example~3 minimizes the peak amplitude relative uncertainty of the $204$~keV peak in the $204+208$~keV Pu doublet, in a $400$-minute measurement using the LANL M400 detector.
The $208$~keV peak is useful in the assay of aged Pu samples, and the $204/208$ ratio in particular can be used to determine the Pu-239/Pu-241 ratio in low-burnup material~\cite[\S 8.3.6]{reilly1991passive}.
Here we fit the peaks with Gaussian models (plus a linear background) since the close spacing of the peaks tends to reduce the observed peak asymmetry and the increased parameter count of the doublet fit makes it challenging to reliably fit two Doniach peaks.
The parameter sweep was the same as in Example~1 except for a reduced range of $\ncomp = 1$--$6$, and ran in ${\sim}2.5$~hours.

\begin{figure}[!htbp]
    \centering
    \includegraphics[width=1.0\linewidth]{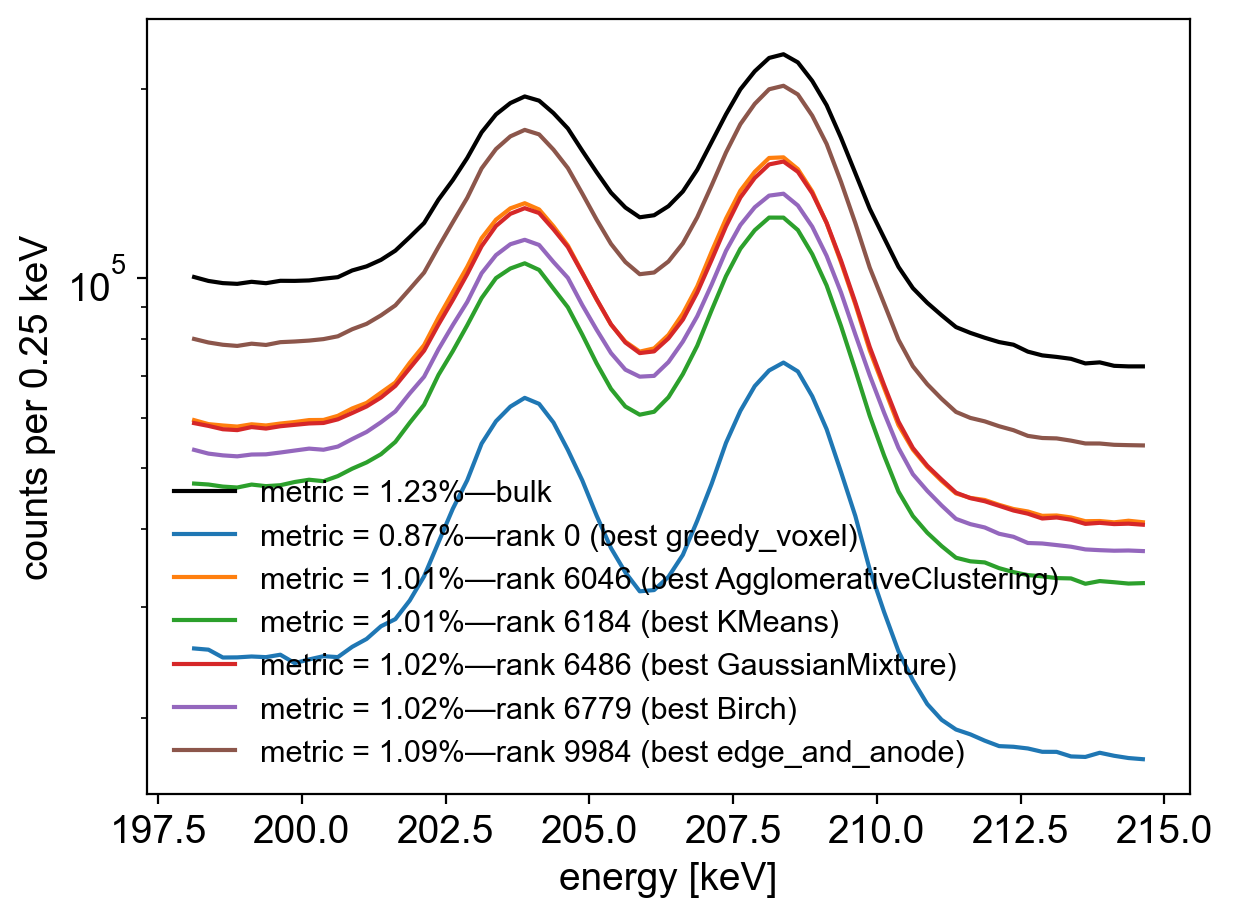}
    \caption{
        Best spectra for select clusterers in Example~3.
    }
    \label{fig:example_gamma}
\end{figure}

Fig.~\ref{fig:example_gamma} shows that the best clustering result (Agglomerative Clustering, $4$~NMF components, $\alpha_W = 0$, $3/4$ clusters retained) improves the Gaussian amplitude relative uncertainty from the bulk value of $1.23\%$ to $1.01\%$.
This $22\%$ relative improvement results from a reduction in relative detector efficiency to $63\%$.
The greedy voxel algorithm however performs the best overall, reducing the metric to $0.87\%$.

\begin{figure*}[!htbp]
    \centering
    \includegraphics[width=0.32\linewidth]{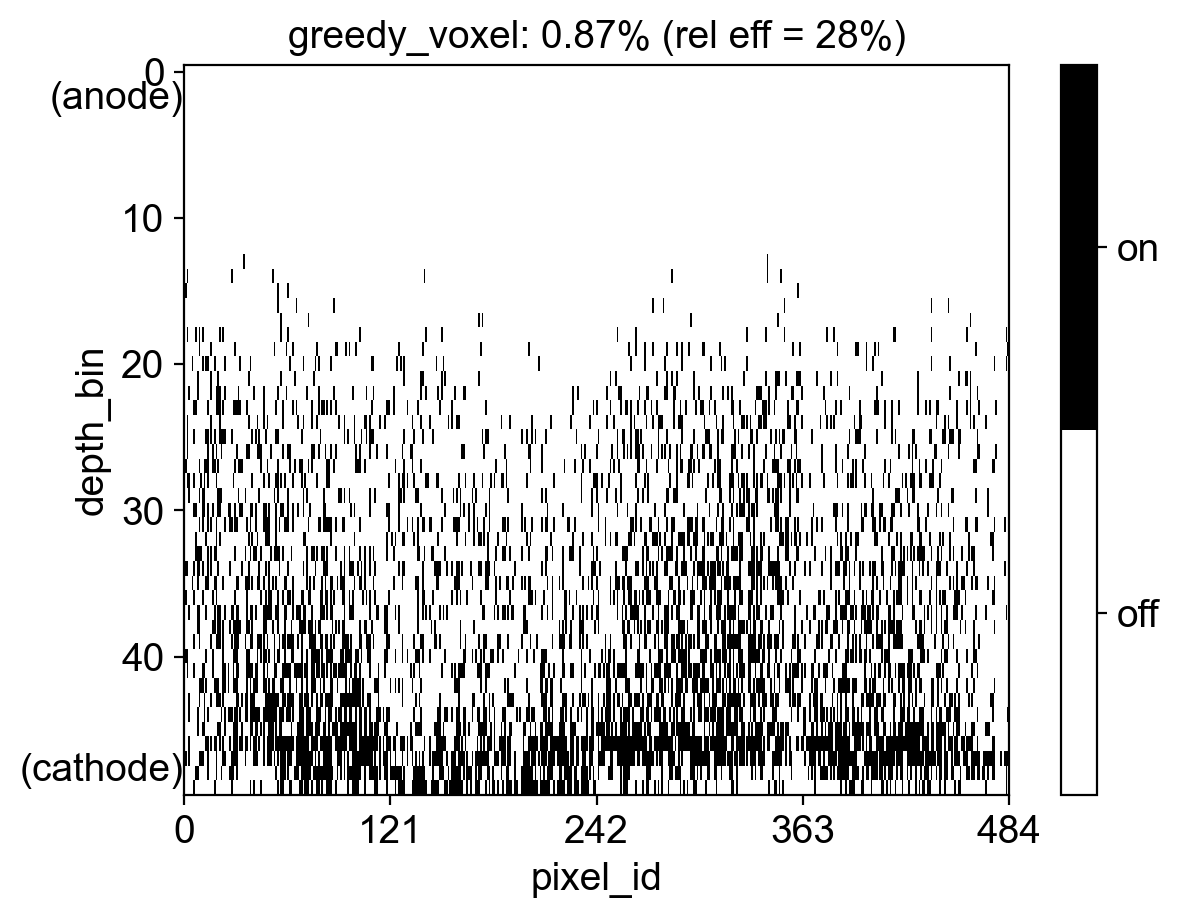}
    \includegraphics[width=0.32\linewidth]{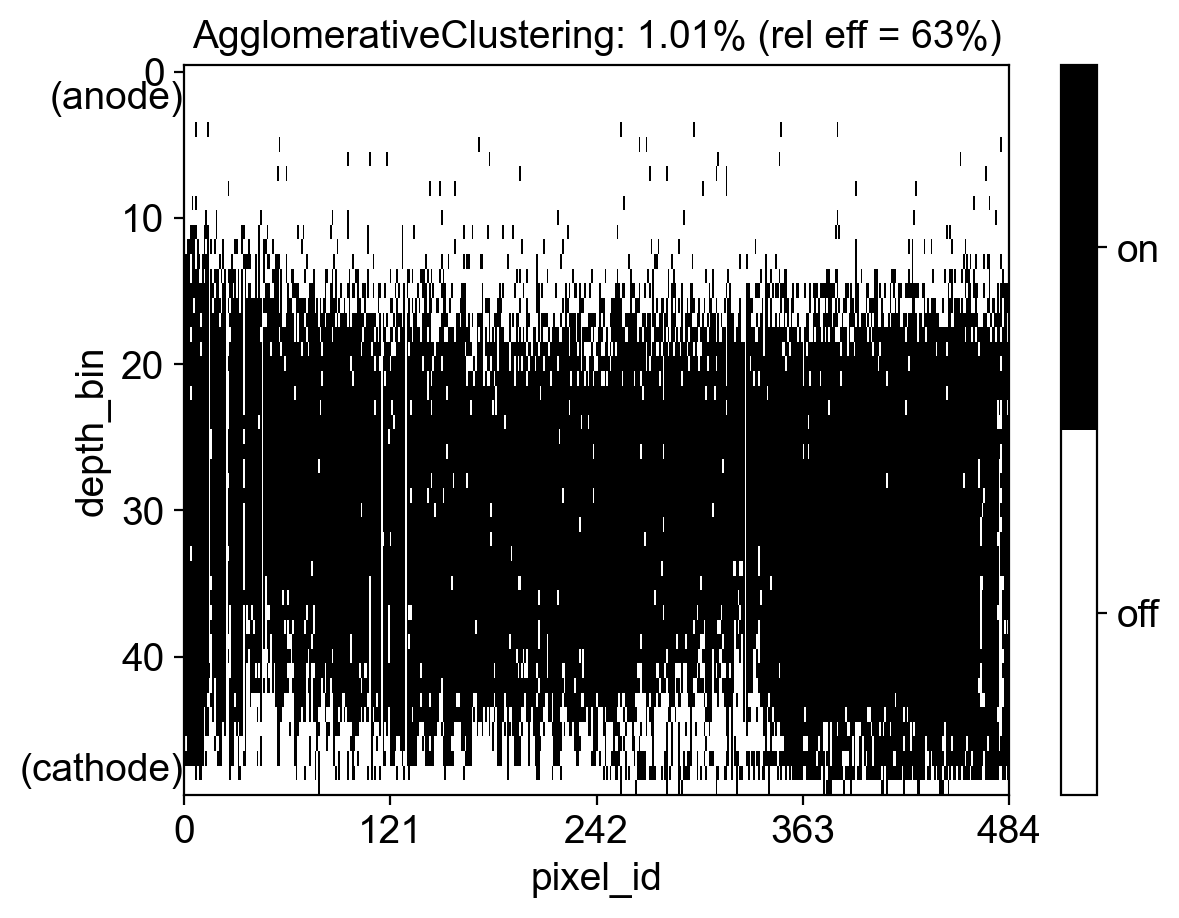}
    \includegraphics[width=0.32\linewidth]{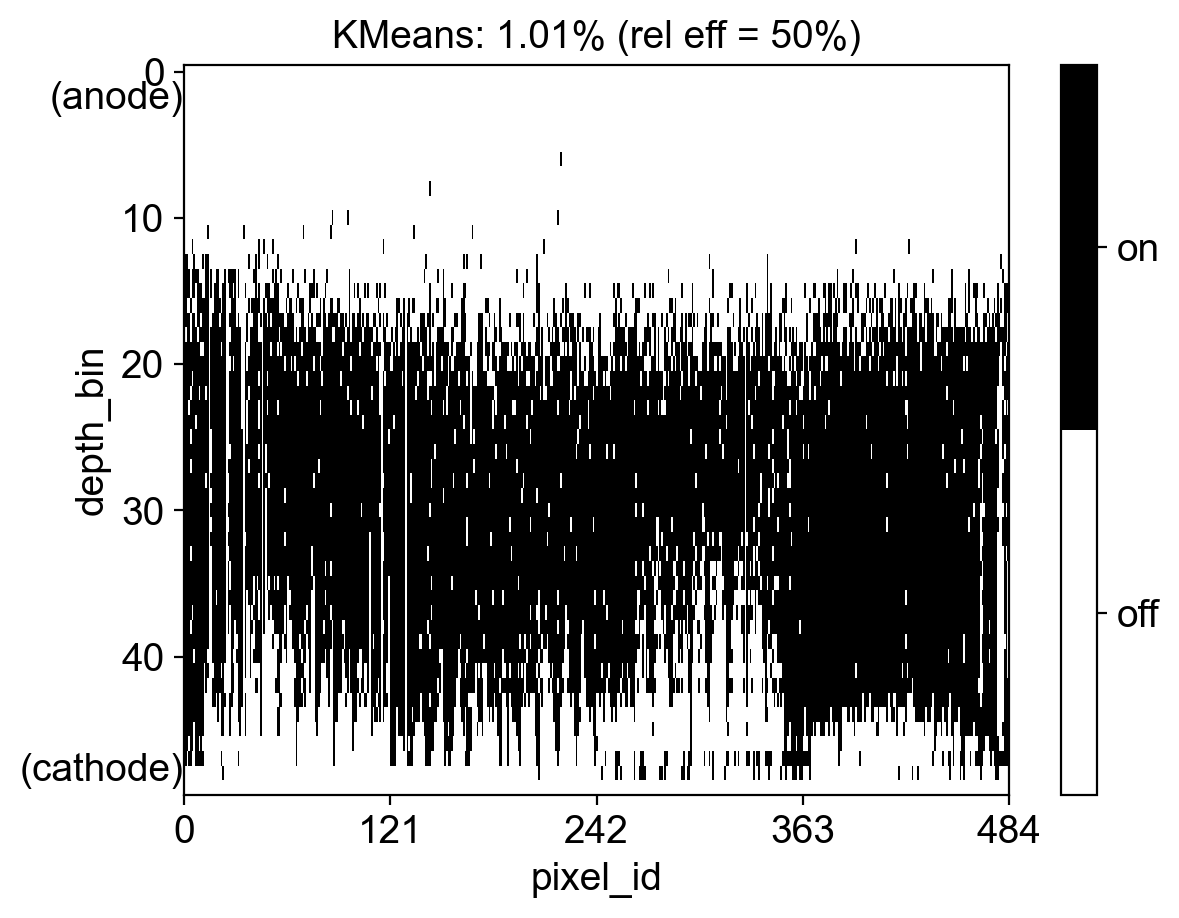}\\
    \includegraphics[width=0.32\linewidth]{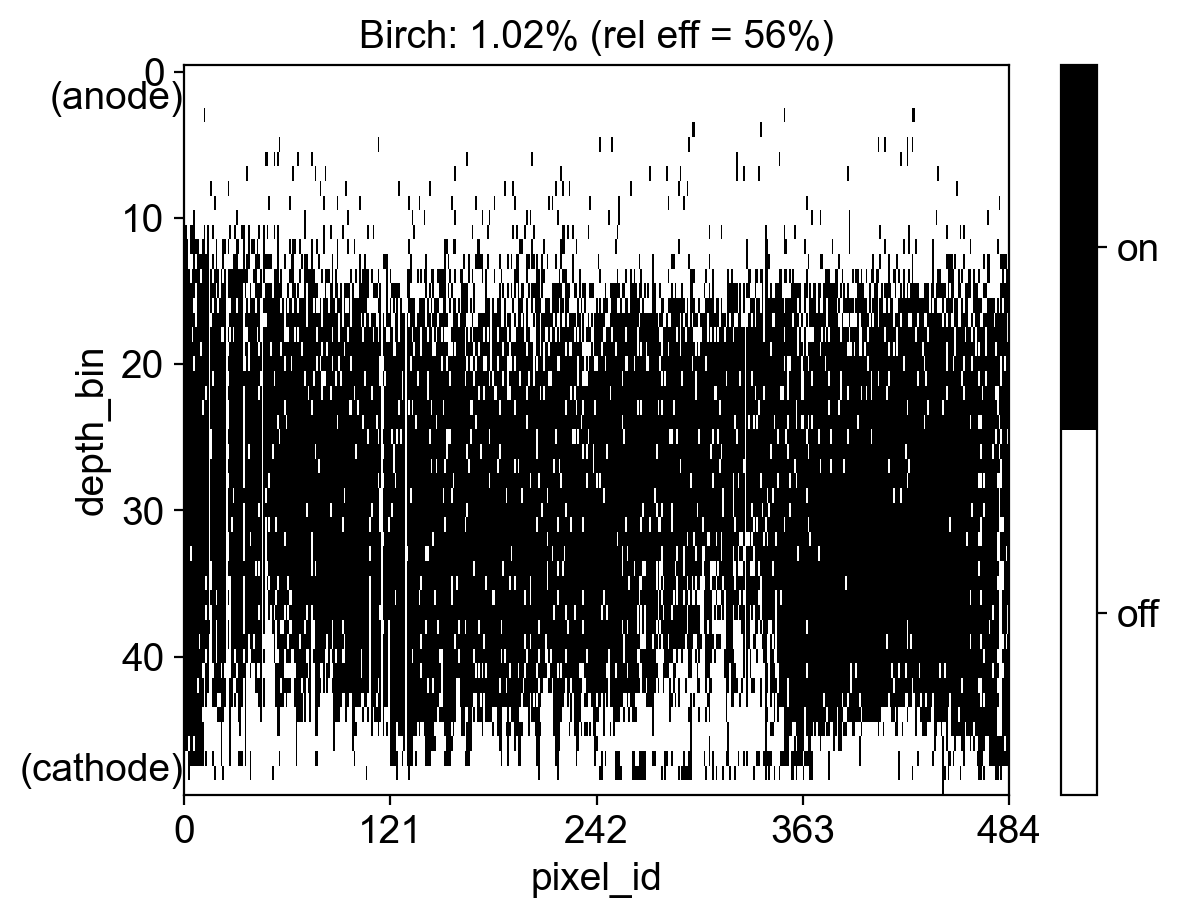}
    \includegraphics[width=0.32\linewidth]{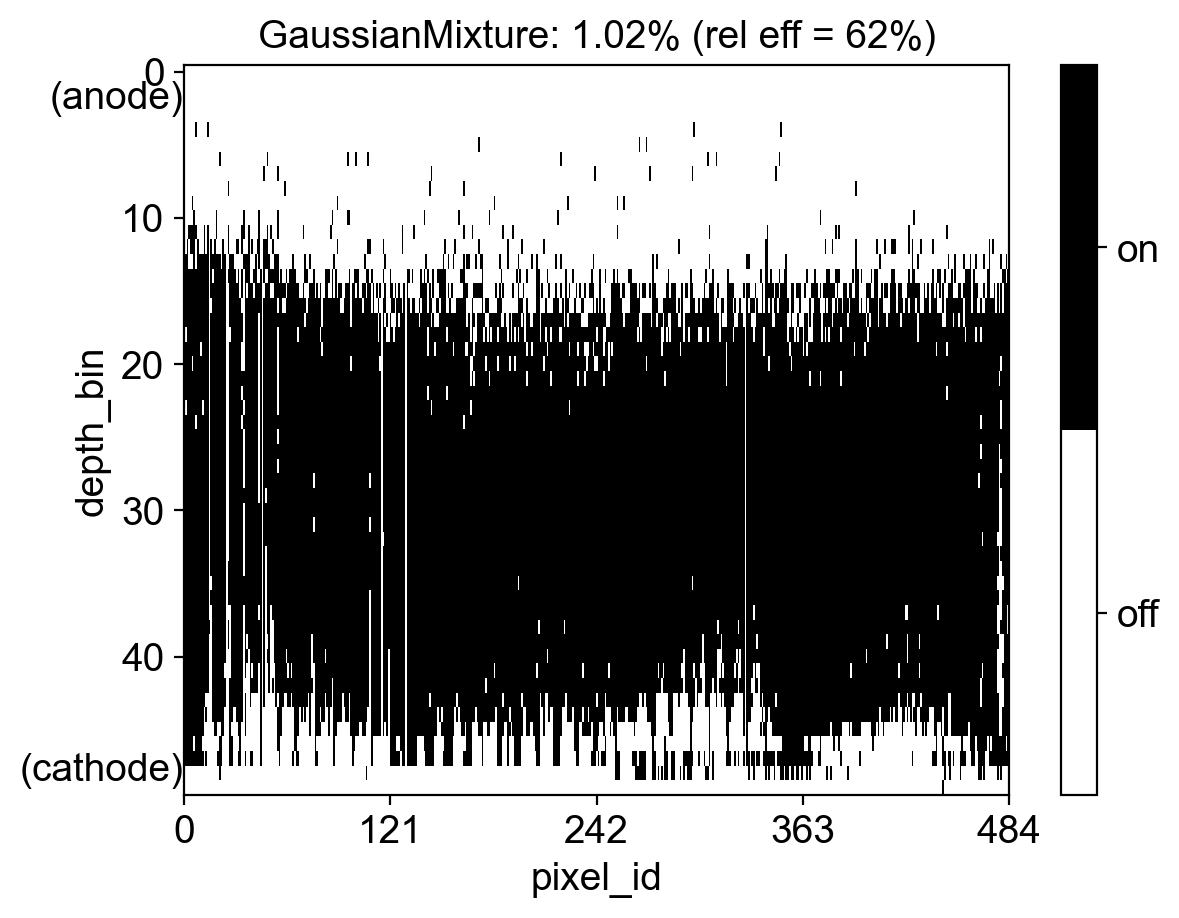}
    \includegraphics[width=0.32\linewidth]{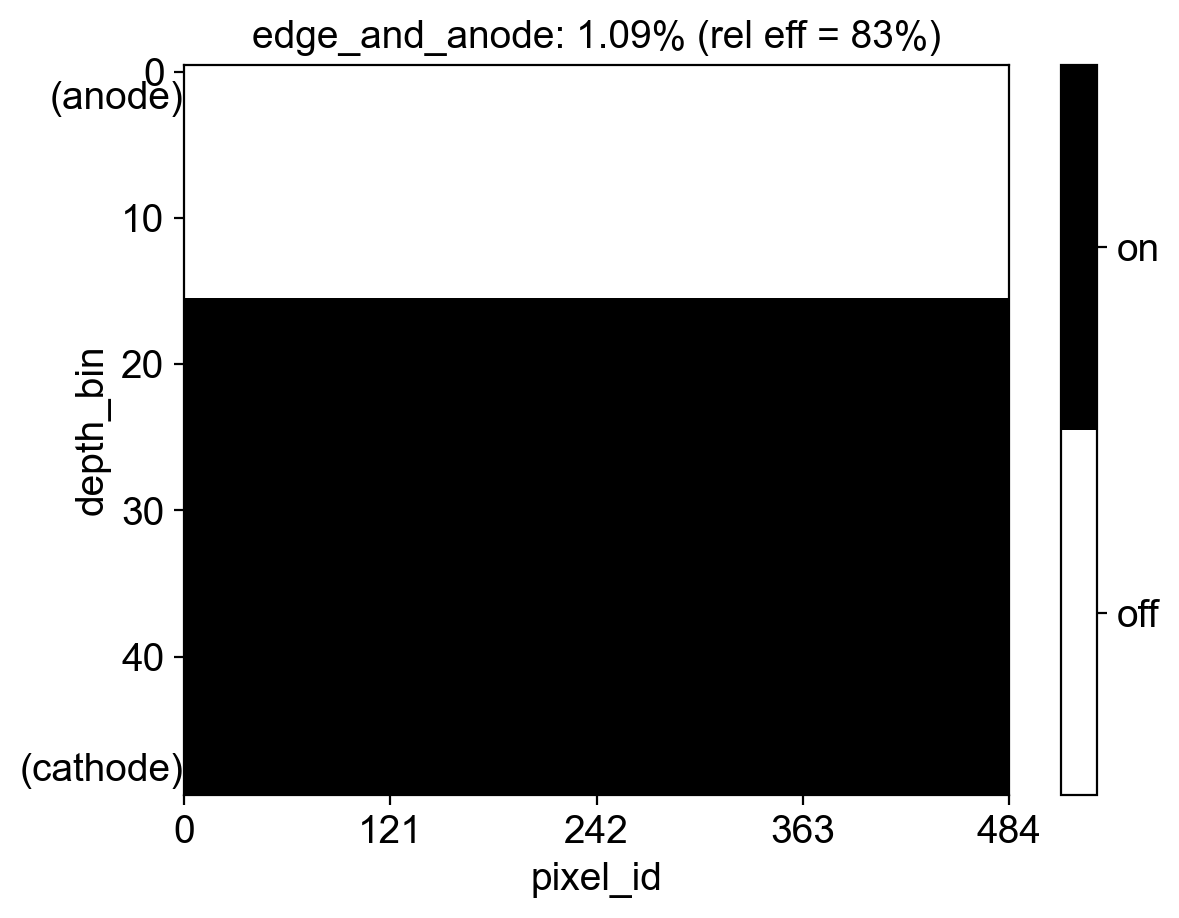}
    \caption{
        Best masks for selected clusterers in Example~3.
    }
    \label{fig:example_gamma_masks}
\end{figure*}

Fig.~\ref{fig:example_gamma_masks} shows the corresponding best active mask for each selected clusterer type in Fig.~\ref{fig:example_gamma}.
All six masks remove most of the quarter of the detector volume closest to the anode.
The ML clusterers additionally remove some voxels near the cathode.
The greedy voxel algorithm follows a similar anode-quarter removal pattern but also removes many more voxels within the the connected regions found by the other clusterers, reducing its relative efficiency to $28\%$.

\begin{figure}[!htbp]
    \centering
    \includegraphics[width=1.0\linewidth]{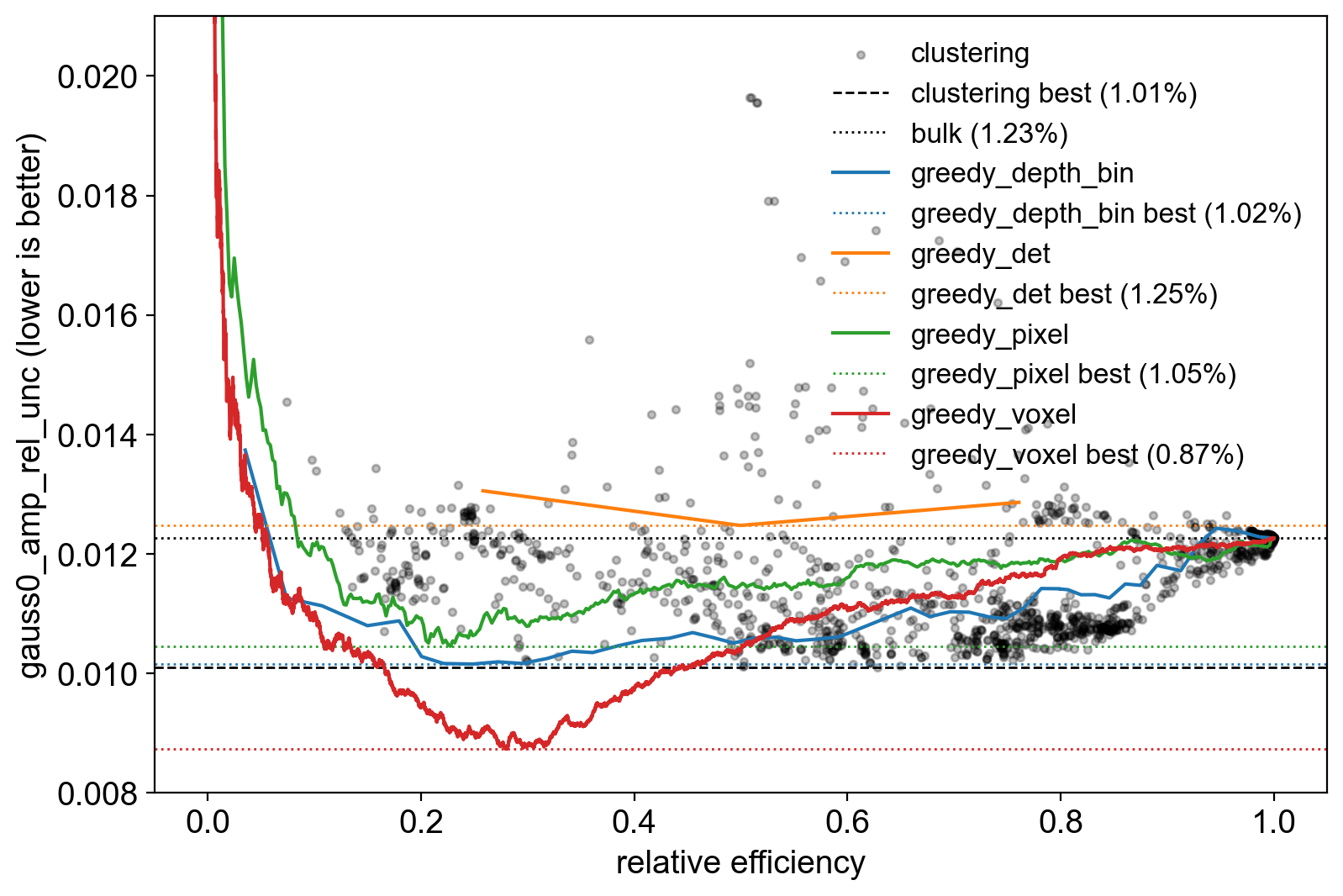}
    \caption{
        Gaussian amplitude relative uncertainty vs.\ relative efficiency for Example~3.
        Points from the random clusterers have been excluded for clarity.
    }
    \label{fig:example_gamma_metric_vs_rel_eff}
\end{figure}

Fig.~\ref{fig:example_gamma_metric_vs_rel_eff} shows the metric vs.\ relative efficiency and bears some similarity to Fig.~\ref{fig:example_beta_metric_vs_rel_eff}.
The greedy voxel, depth bin, and pixel algorithms again start out with high relative uncertainties and reach their minima around efficiencies of $20$--$30\%$ before climbing back up to the bulk metric value.
The greedy detector algorithm performs worse than most clustering models and the greedy pixel algorithm performs slightly better than clustering below efficiencies of ${\sim}50\%$ and worse above.
In contrast, the greedy depth bin and especially voxel algorithms perform substantially better than SPECTRE-ML below ${\sim}50\%$ efficiency, and in fact the greedy voxel algorithm attains the best observed amplitude relative uncertainty of $0.98\%$ at a relative efficiency of $28\%$.

\section{Discussion}\label{sec:discussion}

In Section~\ref{sec:results} we gave four spectral optimization demonstrations for various spectra, detectors, and performance metrics.
Here we provide some additional discussion, including limitations and ongoing/future work.

An important consideration for improving safeguards measurements is the operational ease of finding and applying the optimal voxel mask for a given detector and performance metric.
Thus, while the {\tt spectre-ml} parameter sweeps are certainly more computationally feasible than brute-force optimization over $24\,200$~voxels, the ${\sim}2$~hour runtimes may be inconvenient, especially if a new optimization is desired for every new detector.
In general it is hard to know whether one has tested enough parameter combinations, e.g., whether the $\nclus$ or $\ncomp$ ranges should be expanded, at increased runtime.
The long tail of metric values in the histogram of Fig.~\ref{fig:example_beta}, for example, suggests that better models may be found with more parameter combinations tested.
At the same time, it is hard to determine whether there are any definitive trends in parameters across the various examples shown, e.g., in terms of clustering algorithm, $\nclus$, $\ncomp$, or $\alpha_W$.
The box-and-whisker plot of inverse resolvability vs.\ clusterer in Fig.~\ref{fig:example_alpha}, the plot of best spectrum per clusterer in Fig.~\ref{fig:example_alpha}, and the masks in Fig.~\ref{fig:example_gamma_masks} for instance show that several clusterers are capable of achieving near-optimum results.
In our experience, Agglomerative Clustering often tends to perform well, and it could be useful to discard all other ML clustering algorithms in order to reduce the algorithm runtime.
Here, the black-box nature of the underlying machine learning algorithms is a double-edged sword---while we primarily care about the end optimization metric, better insight into why certain ML parameter combinations perform better or worse could inform future optimization calculations~\cite{bandstra2023explaining}.

Of the greedy algorithms, the greedy depth bin variant in particular is a promising fast approximation to the full SPECTRE-ML parameter sweep, at least at low energies.
In Examples~1a and 2, the greedy depth bin result is similar to the SPECTRE-ML result both in the metric improvement and in the associated efficiency reduction, while in Example~3 the greedy depth bin and voxel algorithms trade off much more relative efficiency to attain similar or better results than SPECTRE-ML.
Thus additional work is needed to quantify whether these models (the greedy models especially) reliably give performance improvements across repeated measurements, and are not just statistical flukes of the training data.
This analysis across repeated measurements could also be expanded to determine whether the algorithms (whether SPECTRE-ML, greedy, or heuristic) generalize across different M400 detectors, or even whether models trained on one detector can improve performance on another.
To this end we are currently investigating performance variations across the six-detector uranium measurement dataset of Ref.~\cite{smith2024summary}, and plan to address these questions in upcoming work.

Also related to generalizability, as demonstrated in Section~\ref{sec:example_beta}, the ML framework is susceptible to specification gaming---specifically, by reducing systematic model fit error---and thus it is vital to carefully define the performance metric to be optimized.
Model error specification gaming could be reduced by pre-processing the training data (e.g., correcting for small pixel-level calibration drifts and possibly even using a fixed centroid parameter), by careful choice of the energy region of interest (to remove contaminant peaks or hard-to-fit backgrounds), or by using more complex but accurate peak models.
To this end, future work will involve replacing our Doniach and Gaussian peak fits with more advanced spectral fitting via the GEM spectroscopy software~\cite{berlizov2022gem, zalavadia2024uranium} that will be used by the IAEA for NDA tasks.

The efficiency reductions in some of the optimized results may, at first glance, appear unacceptably large.
For instance, the best results from both Examples~2 and 3 exhibit efficiency reductions to ${\sim}25\%$ of the bulk, unoptimized detector.
It should be emphasized however that the efficiency loss \textit{alone} is irrelevant as long as the performance metric---which balances efficiency against (typically) resolution---increases.
In general, it is up to the end-user to accurately encode how strongly the efficiency loss should be penalized depending on the application.

All examples in Section~\ref{sec:results} involve the detector being irradiated face-on along its central axis.
While this configuration is appropriate for most NDA measurements, we have not considered measurements with other source-to-detector orientations.
Optimizing a detector for source search or mapping applications, for instance, would likely require training data at the voxel level for multiple directions in $4\pi$, and quite different performance metrics to optimize.

How to extend the optimization to Compton-scattering events that become increasingly important at energies well above the maximum of $1274$~keV here remains an open question.
Directly characterizing and clustering the $6 \times 10^8$ possible ordered voxel \textit{pairs} for $2$-interaction events would become extremely computationally challenging, so alternative approaches would be required.
It would also be interesting to develop and test performance metrics that simultaneously include multiple photopeaks at more disparate energies, such as both the $186$ and $1001$~keV peaks used in the peak ratio method for uranium enrichment assay.

Finally, we have recently learned that the listmode position data from the M400 series encodes information about charge-sharing among detector pixels for each event.
Events with no charge-sharing (single-pixel events) generally have better-resolution spectra than events with charge sharing among $2$ or $3+$ pixels, but removing those events reduces efficiency, very similar to the voxel tradeoff in the present work.
Therefore we are actively working on integrating charge-sharing cuts into our pipelines as another parameter over which to optimize.

\section{Conclusions}
We have introduced a framework for optimizing the spectroscopic performance of detectors with varying performance across their individual readout channels, and applied it to several safeguards-relevant measurements using the H3D M400 gamma spectrometer.
Both the machine learning clustering pipeline and several greedy algorithms were found to improve various performance metrics, including one case in which the relative uncertainty metric was reduced by a factor of $2.9\times$.
The greedy depth bin algorithm at low energies in particular often achieves similar performance improvements as the machine learning pipeline but runs in seconds rather than $2$--$3$~hours.
The various spectra, detectors, and performance metrics used highlight the general nature of the framework.
While the framework generally delivers improved gamma spectra, results will vary depending on the end-user application (i.e., the exact performance metric), and care must be taken to avoid specification gaming.
Future work will examine the generalizability of these results across detectors and improve spectroscopic performance for in-field IAEA NDA measurements.

\section*{Acknowledgments}
The authors thank Duc Vo (Los Alamos National Laboratory) for providing datasets used in this work.
The authors also thank Michael Streicher (H3D, Inc.) and Alain Lebrun, Yannick Dodane, and Andriy Berlizov (International Atomic Energy Agency) for useful discussions.

\bibliographystyle{unsrt}
\bibliography{biblio}

\begin{appendices}

\section{Derivation of the resolvability metric}\label{sec:appendix}

Consider two Gaussian lines, of similar but potentially different strengths $A_0$ and $A_1$.
We wish to quantify how well we can ``detect'' a nonzero fractional difference in their strengths.
The model is
\begin{align}
    f(E | A_0, A_1, \mu_0, \mu_1, \sigma)
        &=
            \frac{
                A_0
            }{
                \sqrt{2 \pi}
                \sigma
            }
            \exp\left(
                -
                \frac{
                    (E - \mu_0)^2
                }{
                    2
                    \sigma^2
                }
            \right)
        \\
        &\ 
            +
            \frac{
                A_1
            }{
                \sqrt{2 \pi}
                \sigma
            }
            \exp\left(
                -
                \frac{
                    (E - \mu_1)^2
                }{
                    2
                    \sigma^2
                }
            \right)
\end{align}
where the peak centroids are $\mu_0 < \mu_1$ and the peak widths~$\sigma$ are assumed to be equal for simplicity.
The observed energy $E$ can be thought of as a random variable where each sample is drawn from the following probability density function:
\begin{align}
    p(E | \alpha_0, \alpha_1, \mu_0, \mu_1, \sigma)
        &=
            \frac{
                \alpha_0
            }{
                \sqrt{2 \pi}
                \sigma
            }
            \exp\left(
                -
                \frac{
                    (E - \mu_0)^2
                }{
                    2
                    \sigma^2
                }
            \right)
        \\
        &\ 
            +
            \frac{
                \alpha_1
            }{
                \sqrt{2 \pi}
                \sigma
            }
            \exp\left(
                -
                \frac{
                    (E - \mu_1)^2
                }{
                    2
                    \sigma^2
                }
            \right)
        ,
\end{align}
where \( \alpha_j \equiv A_j / (A_0 + A_1) \).
Changing variables to the fractional difference \( \delta = (\alpha_1 - \alpha_0)/2 \), \( \bar{\mu} = (\mu_0 + \mu_1) / 2 \), \( \Delta \mu = \mu_1 - \mu_0 \), \( z = \Delta \mu / \sigma \), and \( x = (E - \bar{\mu}) / \sigma \),
\begin{align}
    p(x | \delta, z)
        &=
            \frac{
                1 - 2 \delta
            }{
                2
                \sqrt{2 \pi}
            }
            \exp\left(
                -
                \frac{
                    (x + z/2)^2
                }{
                    2
                }
            \right)
        \\
        &\ 
            +
            \frac{
                1 + 2 \delta
            }{
                2
                \sqrt{2 \pi}
            }
            \exp\left(
                -
                \frac{
                    (x - z/2)^2
                }{
                    2
                }
            \right)
        \\
        &=
            \frac{
                1
            }{
                \sqrt{2 \pi}
            }
            e^{-x^2/2 - z^2/8}
            \left[
                \cosh\left(
                    \frac{x z}{2}
                \right)
                +
                2
                \delta
                \sinh\left(
                    \frac{x z}{2}
                \right)
            \right]
        .
\end{align}

Each sample \( x \) gives us some information about the parameter \( \delta \).
The information per sample can be quantified using the Fisher information:
\begin{align}
    {\cal I}_{\delta\delta}
        &\equiv
            E\left[
                \left(
                    \frac{\partial}{\partial \delta}
                    \log p(x | \delta, z)
                \right)^2
            \right]
        \\
        &=
            \int_{-\infty}^{+\infty}
                \left(
                    \frac{\partial}{\partial \delta}
                    \log p(x | \delta, z)
                \right)^2
                p(x | \delta, z)
                \,
            dx
        \\
        &=
            \int_{-\infty}^{+\infty}
                \frac{
                    \left[
                        \frac{\partial}{\partial \delta}
                        p(x | \delta, z)
                    \right]^2
                }{
                    p(x | \delta, z)
                }
            dx
        \\
        &=
            \frac{
                4
            }{
                \sqrt{2 \pi}
            }
            \int_{-\infty}^{+\infty}
                \frac{
                    e^{-x^2/2 - z^2/8}
                    \sinh^2\left(
                        \frac{x z}{2}
                    \right)
                }{
                    \cosh\left(
                        \frac{x z}{2}
                    \right)
                    +
                    2
                    \delta
                    \sinh\left(
                        \frac{x z}{2}
                    \right)
                }
            dx
        .
\end{align}
We can expand the integrand as a Taylor series around \( z = 0 \) and integrate each term:
\begin{align}
    {\cal I}_{\delta\delta}
        &=
            \frac{
                4
            }{
                \sqrt{2 \pi}
            }
            \int_{-\infty}^{+\infty}
                e^{-x^2/2}
                \left[
                    \frac{1}{4}
                    x^2
                    z^2
                    -
                    \frac{1}{4}
                    \delta\,
                    x^3
                    z^3
                    +
                    {\cal O}(z^4)
                \right]
            dx
        \\
        &=
            z^2
            +
            {\cal O}(z^4)
        .
\end{align}
Information increases in proportion to the number of measurements, so for \( N \) samples of \( x \), the total information to lowest order in \( z \) is
\begin{align}
    {\cal I}_{\delta\delta}
        &\approx
            z^2
            N
        .
\end{align}
The variance of an efficient estimator for $\delta$ is approximately the inverse of the Fisher information:
\begin{align}
    \mathrm{var}[\hat{\delta}]
        &\approx
            z^{-2}
            N^{-1}
        .
\end{align}
So, assuming the number of samples is proportional to the total line strength \( A \), we can construct the expected signal-to-noise ratio of an efficient estimator for \( \delta \) as our resolvability metric:
\begin{align}
    r
        &\equiv
            \frac{
                \delta
            }{
                \sqrt{
                    \mathrm{var}[\hat{\delta}]
                }
            }
        \propto
            \frac{
                \sqrt{A}
            }{
                \sigma
            }
        .
\end{align}

\end{appendices}

\end{document}